\documentclass[preprint2]{aastex62}

\usepackage{cleveref}
\newcommand{\fesclyc}{$f_{esc}^{LyC}$}
\newcommand{\fesclycrel}{$F_{\lambda \rm LyC}/F_{\lambda 1100}$}
\newcommand{\fesclya}{$f_{esc}^{Ly\alpha}$}
\newcommand{\orat}{O$_{32}$}
\newcommand{\nsfr}{$\rho_{\mathrm{SFR}}$}
\newcommand{\xiion}{$\xi_{ion}$}
\newcommand{\sigsfr}{$\Sigma_{SFR}$}

\newcommand{\pubsamp}{\citet{2016Natur.529..178I,2016MNRAS.461.3683I,2018MNRAS.474.4514I,2018MNRAS.478.4851I,2021MNRAS.503.1734I,2019ApJ...885...57W}}

\graphicspath{{./}{./plots/}}

\shorttitle{LzLCS I: New Local LCEs}
\shortauthors{Flury et al.}
\submitjournal{ApJS}
\received{}
\revised{}
\accepted{}

\begin{document}

\title{The Low-Redshift Lyman Continuum Survey I:\\New, Diverse Local Lyman-Continuum Emitters}

\correspondingauthor{Sophia Flury}
\email{sflury@umass.edu}

\author[0000-0002-0159-2613]{Sophia R. Flury}
\affiliation{Department of Astronomy, University of Massachusetts Amherst, Amherst, MA 01002, United States}

\author{Anne E. Jaskot}
%\affiliation{Department of Astronomy, University of Massachusetts Amherst, Amherst, MA 01002, United States}
\affiliation{Department of Astronomy, Williams College, Williamstown, MA 01267, United States}

\author{Harry C. Ferguson}
\affiliation{Space Telescope Science Institute, 3700 San Martin Drive, Baltimore, MD, 21218, USA}

\author{Gabor Worseck}
\affiliation{Institut f\"ur Physik und Astronomie, Universit\"at Potsdam, Karl-Liebknecht-Str. 24/25, D-14476 Potsdam, Germany}

\author{Kirill Makan}
\affiliation{Institut f\"ur Physik und Astronomie, Universit\"at Potsdam, Karl-Liebknecht-Str. 24/25, D-14476 Potsdam, Germany}

\author{John Chisholm}
\affiliation{Department of Astronomy, University of Texas at Austin, Austin, TX 78712, United States}

\author[0000-0001-8419-3062]{Alberto Saldana-Lopez}
\affiliation{Department of Astronomy, University of Geneva, Chemin Pegasi 51, 1290 Versoix, Switzerland}

\author{Daniel Schaerer}
\affiliation{Department of Astronomy, University of Geneva, Chemin Pegasi 51, 1290 Versoix, Switzerland}

\author{Stephan McCandliss}
\affiliation{Department of Physics and Astronomy, Johns Hopkins University, Baltimore, MD 21218, United States}

\author{Bingjie Wang}
\affiliation{Department of Physics and Astronomy, Johns Hopkins University, Baltimore, MD 21218, United States}

\author{N. M. Ford}
\affiliation{Department of Astronomy, Williams College, Williamstown, MA 01267, United States}

\author{Timothy Heckman}
\affiliation{Department of Physics and Astronomy, Johns Hopkins University, Baltimore, MD 21218, United States}

\author{Zhiyuan Ji}
\affiliation{Department of Astronomy, University of Massachusetts Amherst, Amherst, MA 01002, United States}

\author{Mauro Giavalisco}
\affiliation{Department of Astronomy, University of Massachusetts Amherst, Amherst, MA 01002, United States}

\author[0000-0001-5758-1000]{Ricardo Amor\'in}
\affil{Instituto de Investigaci\'on Multidisciplinar en Ciencia y Tecnolog\'ia, Universidad de La Serena, Ra\'ul Bitr\'an 1305, La Serena, Chile}
\affiliation{Departamento de Astronom\'ia, Universidad de La Serena, Av. Juan Cisternas 1200 Norte,  La Serena, Chile}

\author{Hakim Atek}
\affiliation{Institut d’astrophysique de Paris, CNRS UMR7095, Sorbonne Universit\'e, 98bis Boulevard Arago, F-75014 Paris, France}

\author{Jeremy Blaizot}
\affiliation{Univ Lyon, Univ Lyon1, Ens de Lyon, CNRS, Centre de Recherche Astrophysique de Lyon UMR5574, F-69230, Saint-Genis-Laval, France}

\author{Sanchayeeta Borthakur}
\affiliation{School of Earth \& Space Exploration, Arizona State University, Tempe, AZ 85287, USA}

\author{Cody Carr}
\affiliation{Minnesota Institute for Astrophysics, School of Physics and
Astronomy, University of Minnesota, 316 Church str SE, Minneapolis, MN 55455,USA}

\author{Marco Castellano}
\affiliation{INAF, Osservatorio Astronomico di Roma, via Frascati 33, I-00078 Monteporzio Catone, Italy}

\author{Stefano Cristiani}
\affiliation{INAF – Osservatorio Astronomico di Trieste, via G. B. Tiepolo 11, I-34143, Trieste, Italy}

\author{Stephane De Barros}
\affiliation{Department of Astronomy, University of Geneva, Chemin Pegasi 51, 1290 Versoix, Switzerland}

\author{Mark Dickinson}
\affiliation{National Optical-Infrared Astronomy Research Laboratory, Tucson, AZ, USA}

\author{Steven L. Finkelstein}
\affiliation{Department of Astronomy, The University of Texas at Austin, Austin, TX, USA}

\author{Brian Fleming}
\affiliation{Laboratory for Atmospheric and Space Physics, Boulder, Colorado, United States}

\author{Fabio Fontanot}
\affiliation{INAF–Osservatorio Astronomico di Trieste, via G.B. Tiepolo, 11, I-34143 Trieste, Italy}

\author{Thibault Garel}
\affiliation{Department of Astronomy, University of Geneva, Chemin Pegasi 51, 1290 Versoix, Switzerland}

\author{Andrea Grazian}
\affiliation{INAF-Osservatorio Astronomico di Padova, Vicolo dell’Osservatorio 5, I-35122, Padova, Italy}

\author{Matthew Hayes}
\affiliation{The Oskar Klein Centre, Department of Astronomy, Stockholm University, AlbaNova, SE-10691 Stockholm, Sweden}

\author{Alaina Henry}
\affiliation{Space Telescope Science Institute, 3700 San Martin Drive, Baltimore, MD, 21218, USA}

%\author{Juna Kollmeier}
%\affiliation{Observatories of the Carnegie Institution for Science, 813 Santa Barbara Street, Pasadena, CA 91101, USA}

\author{Valentin Mauerhofer}
\affiliation{Department of Astronomy, University of Geneva, Chemin Pegasi 51, 1290 Versoix, Switzerland}

\author{Genoveva Micheva}
\affiliation{Leibniz-Institute for Astrophysics Potsdam, An der Sternwarte 16, 14482 Potsdam, Germany}

\author{M. S. Oey}
\affiliation{Department of Astronomy, University of Michigan, Ann Arbor, MI 48109, USA}

%\author{Ivana Orlitova}

\author{Goran Ostlin}
\affiliation{The Oskar Klein Centre, Department of Astronomy, Stockholm University, AlbaNova, SE-10691 Stockholm, Sweden}

\author{Casey Papovich}
\affiliation{George P. and Cynthia Woods Mitchell Institute for Fundamental Physics and Astronomy, Department of Physics and Astronomy,
Texas A\&M University, College Station, TX, USA}

\author{Laura Pentericci}
\affiliation{INAF, Osservatorio Astronomico di Roma, via Frascati 33, I-00078 Monteporzio Catone, Italy}

\author{Swara Ravindranath}
\affiliation{Space Telescope Science Institute, 3700 San Martin Drive, Baltimore, MD, 21218, USA}

\author{Joakim Rosdahl}
\affiliation{Univ Lyon, Univ Lyon1, Ens de Lyon, CNRS, Centre de Recherche Astrophysique de Lyon UMR5574, F-69230, Saint-Genis-Laval, France}

\author{Michael Rutkowski}
\affiliation{Department of Physics and Astronomy, Minnesota State University, Mankato, MN, 56001, USA}

\author{Paola Santini}
\affiliation{INAF, Osservatorio Astronomico di Roma, via Frascati 33, I-00078 Monteporzio Catone, Italy}

\author{Claudia Scarlata}
\affiliation{Minnesota Institute for Astrophysics, School of Physics and
Astronomy, University of Minnesota, 316 Church str SE, Minneapolis, MN 55455,USA}

\author{Harry Teplitz}
\affiliation{Infrared Processing and Analysis Center, California Institute of Technology, Pasadena, CA 91125, USA}

\author{Trinh Thuan}
\affiliation{Astronomy Department, University of Virginia, Charlottesville, VA 22904, USA}

\author{Maxime Trebitsch}
\affiliation{Astronomy, Kapteyn Astronomical Institute, Landleven 12, 9747 AD Groningen, The Netherlands}

\author{Eros Vanzella}
\affiliation{INAF, Osservatorio Astronomico di Bologna, via Gobetti 93/3 I-40129 Bologna, Italy}

\author{Anne Verhamme}
\affiliation{Department of Astronomy, University of Geneva, Chemin Pegasi 51, 1290 Versoix, Switzerland}
\affiliation{Univ. Lyon, Univ. Lyon 1, ENS de Lyon, CNRS, Centre de Recherche Astrophysique de Lyon UMR5574, 69230 Saint-Genis-Laval, France}

%\author{Isak Wold}
%\affiliation{NASA Goddard Space Flight Center, Greenbelt, MD 20771}

\author{Xinfeng Xu}
\affiliation{Department of Physics and Astronomy, Johns Hopkins University, Baltimore, MD 21218, United States}

\begin{abstract}

The origins of Lyman continuum (LyC) photons responsible for the reionization of the universe are as of yet unknown and highly contested. Detecting LyC photons from the epoch of reionization is not possible due to absorption by the intergalactic medium, which has prompted the development of several indirect diagnostics to infer the rate at which galaxies contribute LyC photons to reionize the universe by studying lower-redshift analogs. We present the Low-redshift Lyman Continuum Survey (LzLCS) comprising measurements made with {\it HST}/COS for a $z=0.2-0.4$ sample of 66 galaxies. After careful processing of the FUV spectra, we obtain a total of 35 Lyman continuum emitters (LCEs) detected with 97.725\% confidence, nearly tripling the number of known local LCEs. We estimate escape fractions from the detected LyC flux and upper limits on the undetected LyC flux, finding a range of LyC escape fractions up to 50\%. Of the 35 LzLCS LCEs, 12 have LyC escape fractions greater than 5\%, more than doubling the number of known local LCEs with cosmologically relevant LyC escape.

\end{abstract}


\section{Introduction}\label{sec:intro}

Numerous observations in the last decade indicate that the universe was reionized by a redshift of $z\approx6$. The Gunn-Peterson effect observed as an absorption trough in the continua of distant quasars { \citep[e.g.,][]{1965ApJ...142.1633G,2001AJ....122.2850B}} and as the absorption of Ly$\alpha$ photons \citep[e.g.,][]{2006AJ....132..117F,2020arXiv200913544Y}, the optical depth of the cosmic microwave background in the form of Thomson scattering \citep[e.g.,][]{2020JCAP...09..005P,2020A&A...641A...6P}, and the reduced transmission of Ly$\alpha$ observed at higher redshifts \citep[e.g.,][]{2018ApJ...856....2M,2020MNRAS.493.3194P} all demonstrate that the intergalactic medium (IGM) transitions from neutral to ionized near this redshift. 

While the state of reionization is clear, the nature of the objects that regulated and dominated this process is not. Dwarf galaxies ({ ${ M_\star\lesssim10^9}$ M$_\odot$}), which have weaker gravitational potentials, are more susceptible to clearing of attenuating material by stellar winds and supernovae (SNe), thereby increasing the fraction of Lyman continuum (LyC) photons which escape their host galaxy \citep[e.g.,][]{2010ApJ...710.1239R,2014MNRAS.442.2560W,2015MNRAS.451.2544P}. However, more massive galaxies have the gas reservoirs necessary for high star formation rate densities and experience less suppressive feedback from stellar winds and supernovae, allowing more stars to form and, ergo, more LyC photon production \citep{2013MNRAS.428.2741W}. 

Which galaxy mass regime dominates reionization is a matter of some contention. {Current empirically-motivated models suggest relatively more luminous (albeit still faint) galaxies are the primary, if not sole, source of escaping LyC photons responsible for reionization \citep{2020ApJ...892..109N}.} Other models predict less luminous galaxies dominate reionization \citep[e.g.,][]{2019ApJ...879...36F}, owing largely to steeper luminosity functions than those adopted by \citet{2020ApJ...892..109N}. Local observations of LyC \citep[e.g.,][]{2018MNRAS.478.4851I} seem to favor the dwarf-galaxy scenario suggested by radiation hydrodynamical simulations \citep[e.g.,][]{2017MNRAS.470..224T} as do some holistic models incorporating massive galaxies, active galactic nuclei (AGN), and dwarf galaxies \citep[e.g.,][]{2019ApJ...879...36F,2020MNRAS.495.3065D}. 

%AGN are largely absent from reionization models \citep[e.g.,][]{2015ApJ...802L..19R} because surveys initially suggested that AGN are rare at $z>6$ \citep[][]{2006AJ....132..117F,2014MNRAS.438.2097F}. More recent efforts suggest this population may be undercounted \citep{2015A&A...578A..83G,2020arXiv200602451G}, making the exclusion of AGN questionable. At this time, any contribution by AGN to reionization remains highly uncertain.
% add in Madau 2015?

Although star-forming galaxies remain the most likely candidates for reionization, their exact contribution is still unknown. One of the least constrained parameters in our understanding of reionization is the so-called escape fraction \fesclyc, the fraction of LyC photons which escape from the host galaxy into the IGM \citep{2001ApJ...546..665S}. As it pertains to reionization, \fesclyc\ relates to the cosmic ionization rate by
\begin{equation}\label{eqn:cosmion}
    \dot{n}_{gal} = f_{esc}^{LyC}\xi_{ion}\rho_{\rm SFR}
\end{equation}
where $\dot{n}_{gal}$ is the emission rate of LyC photons by high-redshift ($z\ge6$) galaxies per unit comoving volume, \xiion\ is the total rate of LyC photons produced within their progenitor galaxies per star formation rate (SFR), and \nsfr\ is the volume density of galaxies per unit SFR. All properties on the right-hand side of Equation \ref{eqn:cosmion} may vary with other galaxy properties such as galaxy mass. Further complicating constraints on $\dot{n}_{gal}$ is the fact that, whereas \nsfr\ can be inferred from observations, \fesclyc\ is degenerate with \xiion\ if measured from Balmer emission lines, which complicates estimating \fesclyc\ from LyC measurements. Some constraint on \xiion, such as H$\alpha$ or [\ion{O}{3}]$\lambda5007$\ \citep[e.g.,][]{2016ApJ...831..176B,2016A&A...591L...8S}, is necessary to break this degeneracy \citep[e.g.,][]{2018ApJ...869..123S}. Values of \fesclyc\ required by reionization models typically span 0.1 \citep{2015ApJ...810...71F,2019ApJ...879...36F} to 0.2 \citep{2015ApJ...802L..19R,2020ApJ...892..109N}.

%\subsection{Previous Lyman Continuum Measurements}

Unfortunately, observational constraints on \fesclyc\ have proven difficult to obtain. Early space-based LyC observations of local galaxies yielded upper limits \citep{1995ApJ...454L..19L,2001A&A...375..805D}, suggesting \fesclyc$\lesssim3$\%. Many putative detections at higher redshifts ($z\sim3$) over the next ten years turned out to be non-LyC contamination from lower-redshift interlopers at small angular separation \citep[e.g.,][]{2012ApJ...751...70V,2015ApJ...810..107M,2015ApJ...804...17S}. Another difficulty stems from IGM attenuation at higher redshifts, which (i) makes LyC measurements at or beyond reionization ($z\gtrsim6$) impossible because all the LyC photons are absorbed \citep[e.g.,][]{2021arXiv210316610B}; and (ii) makes imaging detections of the LyC at moderately high redshifts ($z=2-3$) complicated because the Lyman series and LyC attenuation are uncertain along any particular line of sight \citep[e.g.,][]{2014MNRAS.442.1805I,2018ApJ...869..123S}. Moreover, the LyC is too faint at $z\gtrsim4$ to be detected with even the largest contemporary ground-based telescopes.

Despite these complications, the past several years have enjoyed an explosion of LyC detections from a few upper limits to a few tens of significant measurements. Observations of local ($z<0.4$) galaxies with the {\it Hubble Space Telescope} ({\it HST}) Cosmic Origins Spectrograph \citep[COS;][]{2012ApJ...744...60G} have yielded $\gtrsim16$ LyC detections \citep[][]{2013A&A...553A.106L,2014Sci...346..216B,2016ApJ...823...64L,2016Natur.529..178I,2016MNRAS.461.3683I,2018MNRAS.474.4514I,2018MNRAS.478.4851I,2019ApJ...885...57W,2021MNRAS.503.1734I}. These measurements are not without difficulty: scattered telluric light can contaminate the LyC (cf. \citealp{2017A&A...605A..67C} regarding \citealp{2016ApJ...823...64L}) and even masquerade as LyC (cf. \citealp{2016MNRAS.461.3683I} regarding \emph{FUSE} observations by \citealp{2013A&A...553A.106L}).

Of the best measurements of the LyC from local galaxies, the Green Peas (GPs), which in many ways resemble galaxies in early cosmological epochs \citep[e.g.,][]{2009MNRAS.399.1191C,2010ApJ...715L.128A,2012ApJ...749..185A,2013ApJ...766...91J,2016A&A...591L...8S}, exhibit \fesclyc\ which can exceed 20\%\  \citep[e.g.,][]{2018MNRAS.478.4851I}. Ground-based observations have made significant headway at higher redshifts ($z\sim3$) with careful measurements of \fesclyc\ for 16 galaxies \citep{2018ApJ...869..123S,2018MNRAS.476L..15V}. Further {\it HST} observations have contributed to the $z\sim3$ measurements of \fesclyc\ by adding $\gtrsim10$ LyC detections \citep{2015ApJ...810..107M,2016A&A...585A..51D,2016ApJ...825...41V,2017ApJ...837L..12B,2017MNRAS.465..302M,2019ApJ...878...87F,2019Sci...366..738R,2020ApJ...888..109J}. 

%\subsection{Indicators of LyC Escape}

The search for Lyman continuum emitters (LCEs, sometimes referred to as Lyman continuum ``leakers'', e.g., \citealp{2013A&A...554A..38B}) at high redshift relies on indirect indicators of the physical mechanisms involved in LyC escape. The [\ion{O}{3}]$\lambda$5007/[\ion{O}{2}]$\lambda\lambda$3726,29 ($O_{32}$) emission line flux ratio, { thought to be a proxy for optical depth} in extreme GP galaxies \citep[e.g.,][]{2013ApJ...766...91J,2014MNRAS.442..900N}, has successfully been used to select LCE candidates for two {\it HST}/COS observing programs \citep{2016Natur.529..178I,2016MNRAS.461.3683I,2018MNRAS.474.4514I,2018MNRAS.478.4851I}. However, a third observing program using \orat\ as a selection criterion, \citet{2021MNRAS.503.1734I}, did not reproduce this success. Star formation rate surface density (\sigsfr) can gauge the role of stellar feedback in facilitating LyC escape \citep[e.g.,][]{2001ApJ...558...56H,2002MNRAS.337.1299C}. Indeed, cosmological simulations predict that \sigsfr\ correlates with \fesclyc\ where $\Sigma_{SFR}>0.1$ M$_\odot$ yr$^{-1}$ kpc$^{-2}$ corresponds to cosmologically relevant values of \fesclyc$\gtrsim5$\%\ \citep[e.g.,][]{2017MNRAS.468.2176S,2020ApJ...892..109N}. Extinction and starburst age determine the slope (measured by the spectral index assuming $f_\lambda\propto\lambda^\beta$) of the non-ionizing UV continuum. Values of $\beta\lesssim-2$ indicate young ($<$30 Myr), unextinguished stellar populations from which LyC photons could escape \citep[e.g.,][]{2013ApJ...777...39Z,2017ApJ...836...78Z}. Together, \orat, \sigsfr, and $\beta$ can serve as holistic selection criteria for LCE candidates, {although these properties may select galaxies with intrinsically strong LyC flux rather than galaxies with high \fesclyc.}

This paper presents the Low-Redshift Lyman Continuum Survey (LzLCS). We have assembled a sample of 66 star-forming galaxies from the Sloan Digital Sky Survey \citep[SDSS,][]{2000AJ....120.1579Y} and the {\it Galaxy Evolution Explorer} \citep[{\it GALEX},][]{2003SPIE.4854..336M} that reside nearby ($z\sim0.3$) and are considered likely candidates for LyC escape according to the above criteria (\S\ref{sec:sample}). We have observed each of these 66 galaxies with {\it HST}/COS to measure the LyC (\S\ref{sec:obs}-\ref{sec:lyc_meas}). From the {\it HST}/COS and SDSS photometry and spectra, we derive physical and observational properties to characterize these 66 galaxies and compare them to a set of LCEs from the literature which have been previously observed with {\it HST}/COS (\S\ref{sec:global_props}). From our measurements of the LyC flux, we provide estimates of \fesclyc\ for our sample (\S\ref{sec:fesc}). Companion papers present our initial tests of indirect LyC diagnostics \citep[Flury et al. submitted, Salda\~na-Lopez et al. submitted,][]{2021arXiv210403432W}. Throughout this paper, we assume $H_0=70$ km s$^{-1}$\ Mpc$^{-1}$, $\Omega_m=0.3$, and $\Omega_\Lambda=0.7$.

\section{Sample Definition}\label{sec:sample}

We define a sample of 66 LCE candidates, the LzLCS, to investigate the properties, physical mechanisms, and diagnostics associated with LyC escape. To begin, we search for star-forming galaxies in the SDSS Data Release 15 \citep{2017AJ....154...28B} using either the tabulated emission line fluxes or our own measurements made following \citet{2019ApJ...885...96J}. We limit our sample to star-forming galaxies by using the BPT diagram \citep{1981PASP...93....5B} to exclude AGN and composite systems. Then, we match these objects with photometry from the {\it GALEX} data archive. We select galaxies which are relatively nearby ($z\sim0.3$) so that the rest-frame LyC can be readily observed by {\it HST}/COS with the G140L grating while reaching the sensitivity required to detect \fesclyc$\sim5$\% at S/N$>5$ for each object. The COS throughput imposes a redshift limit of $z\gtrsim0.22$ to this detection goal because its sensitivity declines by roughly two orders of magnitude for wavelengths below 1100~\AA\ \citep{2012ApJ...744...60G}.

%The \fesclyc\ diagnostics discussed in \S\ref{sec:intro_diag} are under-explored. 
From the SDSS-{\it GALEX} star-forming galaxies, we select objects to evenly sample the \orat, $\beta$, and \sigsfr\ methods of inferring \fesclyc\ with $\sim50$ objects each across a range of the relevant parameter space.% A sample of this size will improve our ability to infer \fesclyc\ to an accuracy of 15-25\% in each diagnostic by sampling a sufficient number of possible orientations of optically thin cavities \citep[see][]{2015ApJ...801L..25C}. 
{The LzLCS is thus designed to span a far wider range of relevant parameter spaces than previous investigations, allowing us to determine whether LCEs are a heterogeneous or homogeneous population of galaxies.}

\begin{figure}
    \centering
    \includegraphics[width=\columnwidth,trim={5.5in 0.25in 0.5in 1.2in}, clip]{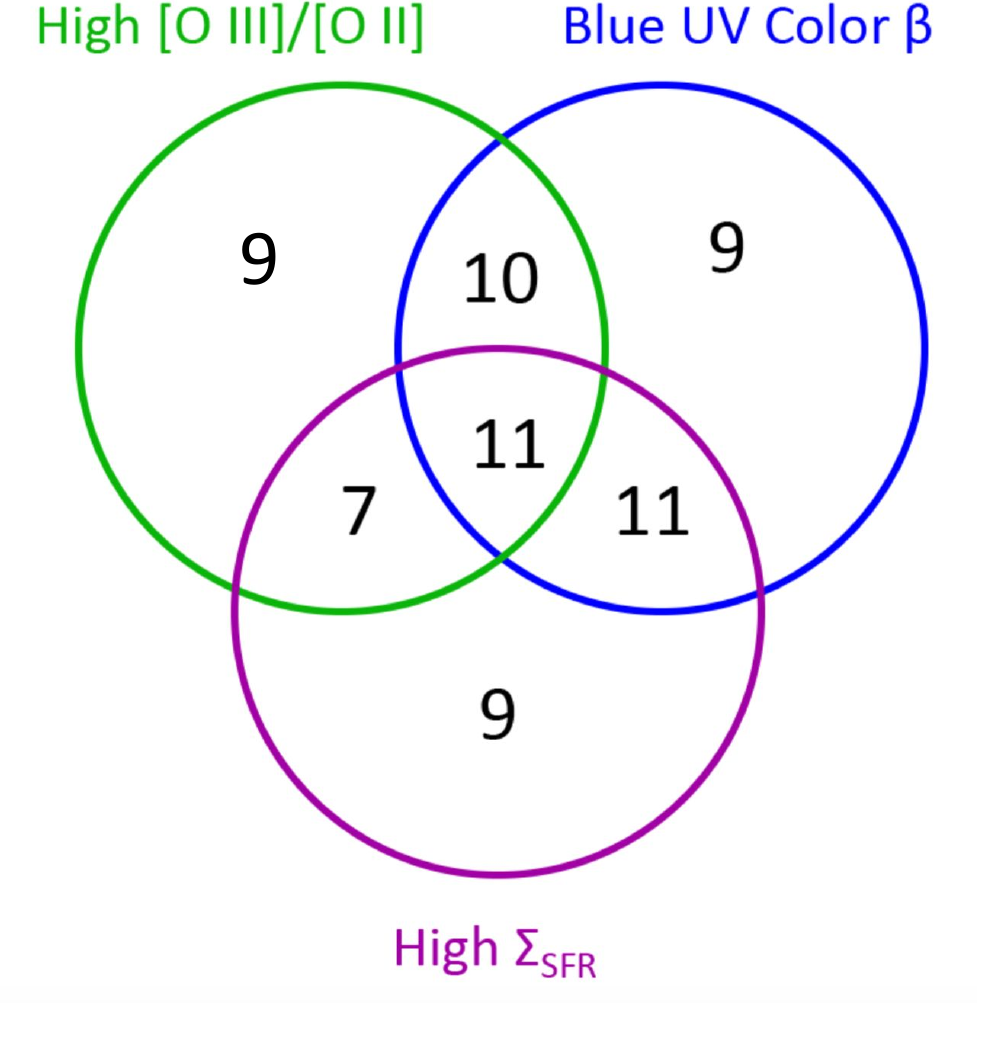}
    %,trim={5.5in 0.25in 0.5in 1.2in}, clip
    \caption{Venn diagram showing the overlap of the three selection criteria for the LzLCS: \orat=[\ion{O}{3}]/[\ion{O}{2}]\ flux ratio (top left), UV slope parameterized by spectral index $\beta$ (top right), and star formation rate surface density \sigsfr\ (bottom center).
    \label{fig:venn}}
\end{figure}

We include 37 objects from the low-redshift SDSS/{\it GALEX} with \orat$>3$, bringing the number of objects with high \orat\ to 50 when combined with previous studies. We include an additional 29 galaxies with high star formation rate surface densities (lower limit of $\Sigma_{SFR}>0.1$ M$_\sun$ yr$^{-1}$ kpc$^{-2}$ estimated from the dust-corrected {\it GALEX} FUV magnitude assuming $A_{FUV}\sim12\times E(B-V)$ and SDSS $u$-band half-light radius) and/or blue UV continua (power law index $\beta<-2$ estimated from the {\it GALEX} photometry). We chose galaxies with various combinations of these criteria to improve the chances of targeting true LCEs. As a result, at least 37 galaxies in the total sample satisfy each criterion with 11 galaxies satisfying all three (Fig \ref{fig:venn}).

\section{Observations with {\it HST}/COS}\label{sec:obs}

\begin{figure*}[!t]
    \centering

    \includegraphics[width=0.32\textwidth,trim={1.5cm 1cm 1.5cm 0cm},clip]{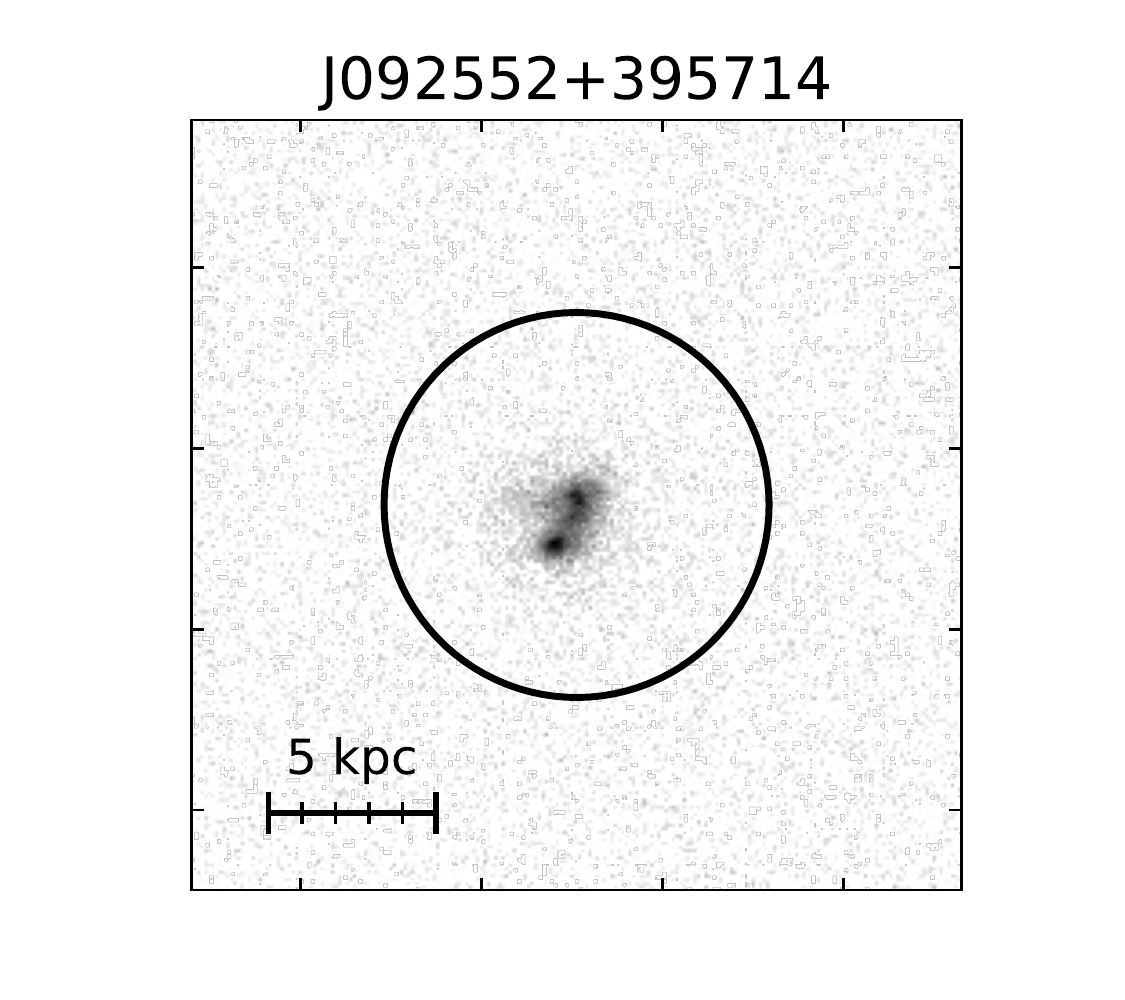}
    \includegraphics[width=0.32\textwidth,trim={1.5cm 1cm 1.5cm 0cm},clip]{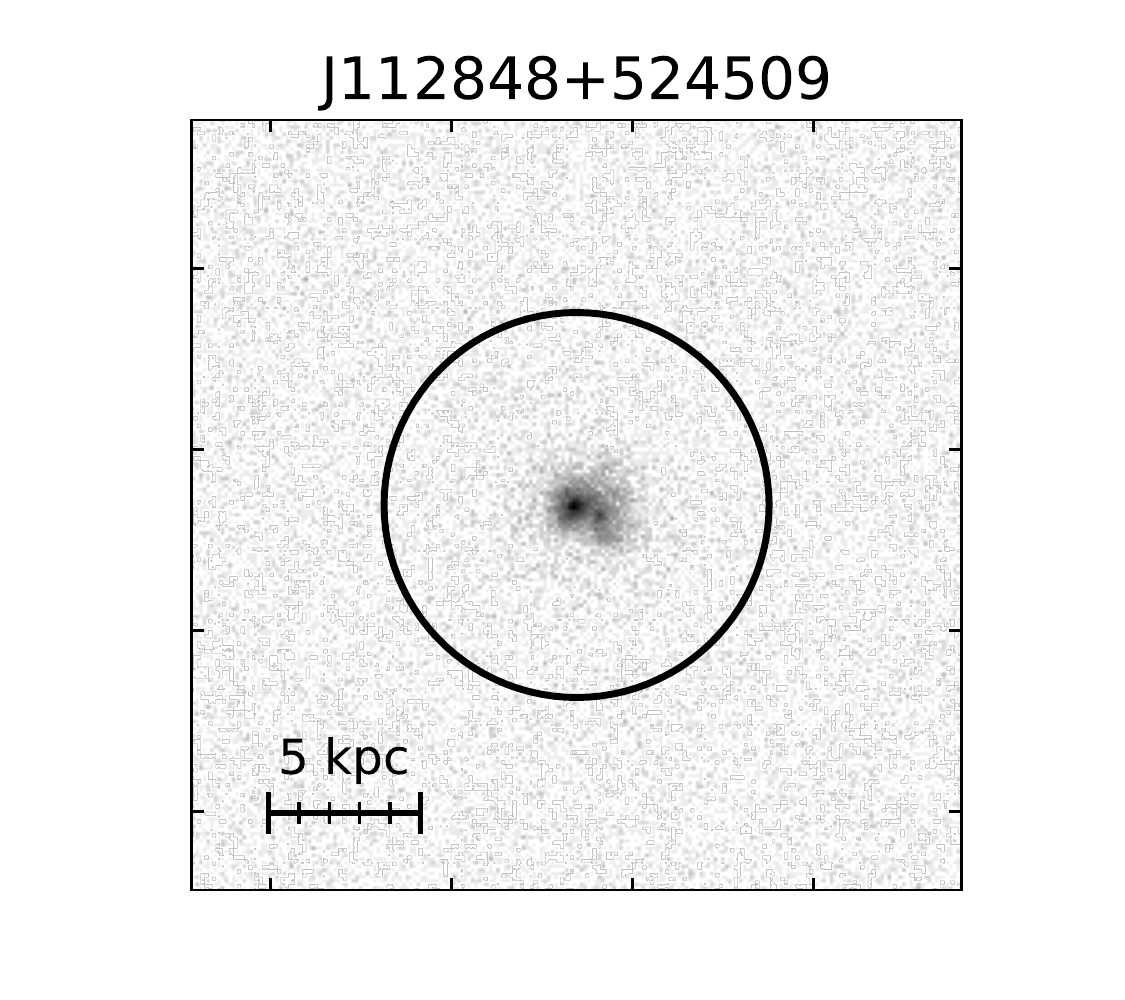}
    \includegraphics[width=0.32\textwidth,trim={1.5cm 1cm 1.5cm 0cm},clip]{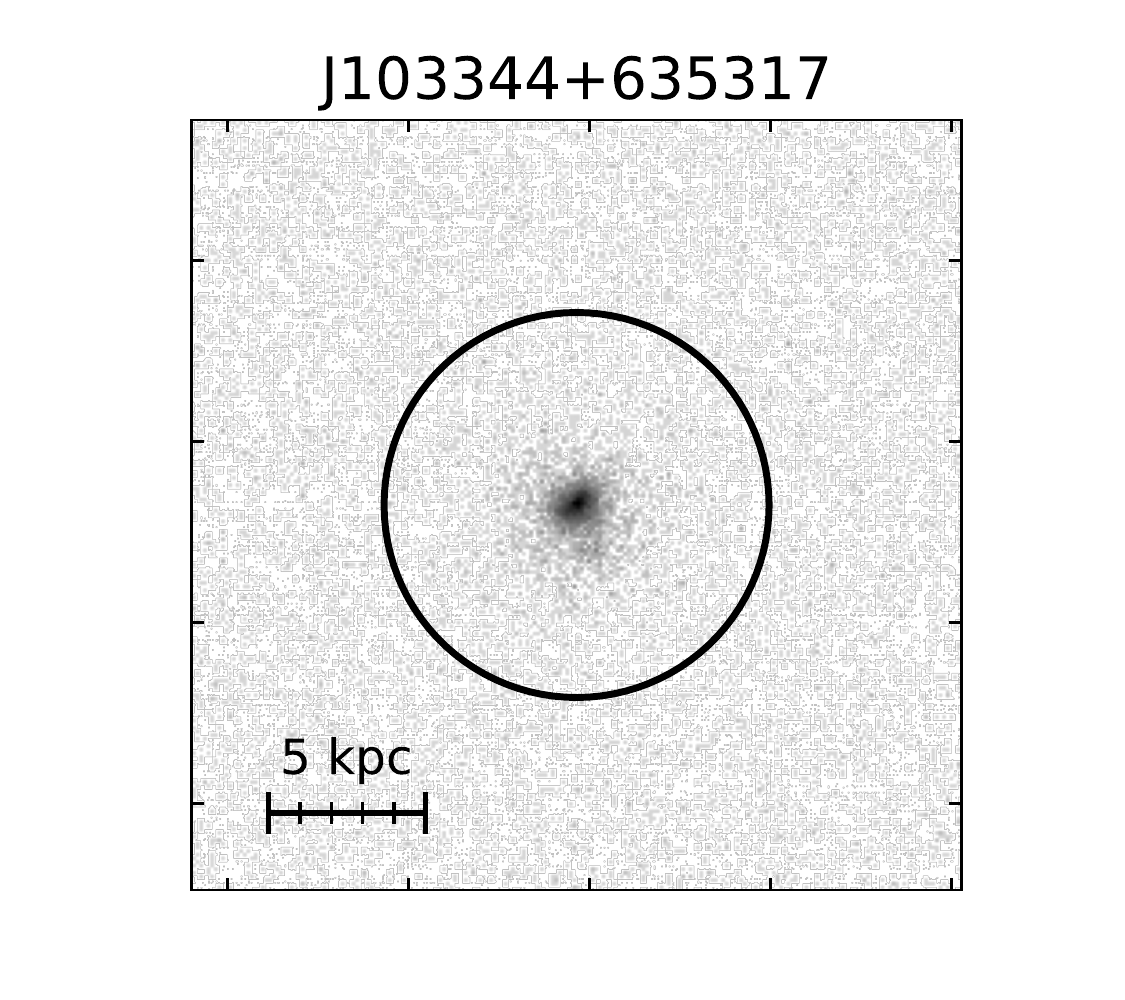}

\caption{Example log-scaled acquisition images of a non-emitter ({left}), marginally detected LCE ({center}) and well detected LCE ({right}) from the LzLCS. Circle indicates the 2.5\arcsec\ COS spectroscopic aperture. Bar in the lower left indicates a physical scale of 5 kpc at the target's redshift with 1 kpc ticks. Acquisition image thumbnails for the full sample are available online.\label{fig:cos_aqims}}

\end{figure*}

For the sample of 66 candidate LyC-leaking galaxies, we obtained 134 orbits of {\it HST}/COS spectroscopy under observing program GO 15626 (Cycle 26, P.I. Jaskot). COS acquires each object via NUV imaging and centered its 2.5\arcsec\ diameter spectroscopic aperture on the peak NUV flux. We used the G140L grating at 800 \AA\ in COS Lifetime Position 4, covering a wavelength range of $800-1950$ \AA\ with { a resolution of $R\sim1050$ at 1100 \AA\ \citep[cf.][]{2021MNRAS.503.1734I}.} We show a log of the observations in Table \ref{tab:obs_log} { and example acquisition images in Figure \ref{fig:cos_aqims}}.

\begin{deluxetable*}{cccccccc}
\tablecaption{{\it HST}/COS observation log for the LzLCS. Also indicated are the average number of dark current observations used to model the background dark current in each visit. A full version of this table is available online. \label{tab:obs_log}}% A full version of this table is available online.}
\tablewidth{\textwidth}
\tablehead{
\colhead{Object} & \colhead{RA (deg)} & \colhead{Dec (deg)} & \colhead{$z$} & \colhead{Visit} & \colhead{Date} & \colhead{Exp Time (s)} & \colhead{N Darks}
}
\startdata
J003601+003307 & ~~9.002641 &~~0.552006 & 0.3479 &ldxe08 & 2019-09-25 &~3980.384 &13 \\
J003601+003307 & ~~9.002641 &~~0.552006 & 0.3479 &ldxew9 & 2019-12-08 &~3980.672 &~9 \\
J004743+015440 & ~11.928487 &~~1.911086 & 0.3535 &ldxe42 & 2019-07-29 &~1495.904 &~9 \\
J011309+000223 & ~18.286905 &~~0.039839 & 0.3062 &ldxez5 & 2019-08-03 &~1363.904 &13 \\
J012217+052044 & ~20.569425 &~~5.345561 & 0.3656 &ldxe05 & 2019-09-20 &~3780.320 &15
\enddata
\end{deluxetable*}

Following previous works \citep{2016ApJ...825..144W,2018MNRAS.478.4851I,2019ApJ...885...57W,2021MNRAS.503.1734I}, we process the raw {\it HST}/COS spectra using a combination of standard and custom software { to best model the background and optimize measurement of the LyC}. The COS detector measures the pulse height amplitude (PHA) of charge produced by an amplifying microchannel plate. Dark current and location-dependent geomagnetic activity like the South Atlantic Anomaly can trigger the COS detector, resulting in spurious background events with a PHA distribution extending beyond PHAs of science events \citep{2016ApJ...825..144W,2016MNRAS.461.3683I}. Before processing the spectra, we screen PHAs to include only values within the $1$\textendash$12$ range for Lifetime Position 4 to mitigate dark current and other without excluding science events. We reduce the spectra using the {\sc calcos} pipeline (v3.3.9) to perform flat-fielding, dead time and stim pulse corrections, and wavelength and flux calibrations. For the extraction, we draw a rectangular aperture 25 pixels wide along the cross-dispersion axis (hatched region in Figure \ref{fig:spec2d}), a range comparable to the 95th percentiles of the total starlight continuum profile after excluding geocoronal Ly$\alpha$ emission.

With the custom software {\sc FaintCOS} \citep{2016ApJ...825..144W,2020arXiv201207876M}\footnote{https://github.com/kimakan/FaintCOS}, we estimate the dark current and scattered geocoronal Ly$\alpha$ background and co-add individual exposures to improve signal-to-noise ratios while preserving Poisson counts. For each observation, we compare the PHA cumulative distribution function of each dark obtained from a $\pm$1 month window, selecting those which match the PHA cumulative distribution function of the background in the science image by means of a Kolmogorov-Smirnoff test ($D<0.03$) to ensure the solar and geomagnetic conditions during the recording of the darks are comparable to those of the science image \citep{2016ApJ...825..144W,2018MNRAS.478.4851I}. We show the number of dark observations selected in this manner for each visit in Table \ref{tab:obs_log}. To evaluate the success of the dark model, we compare the spatial variations of the dark model to the background of the science image and find the two consistently agree (see grey line in Figure \ref{fig:spec2d}). We scale the \citet{2016ApJ...825..144W} model for scattered light background by the peak telluric Ly$\alpha$ counts in each extraction. { We co-add the spectra and background models for each object and bin from the over-sampled detector dispersion of 0.0803~\AA\ to 0.5621~\AA\ in order to Nyquist-sample the G140L resolution of $\sim1.1$~\AA.} {In cases where the gross counts are less than 100 counts above the background, we determine the Neyman-Pearson 1$\sigma$ confidence intervals in the flux measurements following \citet{1998PhRvD..57.3873F}. Otherwise, we sample variates from the Poisson distributions of the background and science spectra to compute the $1$-$\sigma$ confidence in the flux in each pixel. Finally, we correct each spectrum for Milky Way extinction using Galactic $E(B-V)$ estimates from the dust maps by \citet{2018MNRAS.478..651G} and the \citet{1999PASP..111...63F} extinction law. To determine the final uncertainty in the corrected flux in each pixel, we sample flux measurements and the \citet{2018MNRAS.478..651G} reddening $10^4$ times.}

\section{Measuring Lyman Continuum}\label{sec:lyc_meas}

We measure the LyC flux in a rest-frame 20~\AA\ window as close as possible to $\lambda_{rest}=900$~\AA\ while avoiding wavelengths above $\lambda_{obs}=1180$~\AA\ to minimize telluric contamination, rounding down to the nearest 10 \AA\ in the rest frame. This constraint ensures as uniform a measurement of the LyC as possible across the entire sample while simultaneously preventing any contamination by the Ly$\alpha$ and \ion{N}{1} $\lambda1200$ geocoronal emission lines or by non-LyC starlight introduced into the 900-912~\AA\ range by dispersion through the {\it HST}/COS optics. To evaluate any unresolved geocoronal contamination, we compare the count rate during orbital night to that of the total visit and find good agreement between the two, indicating no significant contributions of telluric radiation to the LyC. Moreover, the scattered light model by \citet{2016ApJ...825..144W} rectifies minor discrepancies between the two count rates, indicating that our treatment of the background light is appropriate.
%compute the dark-subtracted count rate variability in the LyC window over each observation and find that {only one detection, with probability $P(>N|B)=0.0145$ that the measurement is drawn from the distribution of background counts (i.e., a marginal detection), contains significant fluctuations attributable to telluric radiation. Visual inspection of the target spectrum suggests the scattered geocoronal light model sufficiently accounts for this contamination.}

\begin{figure*}[!t]
    \centering

    \includegraphics[width=0.32\textwidth]{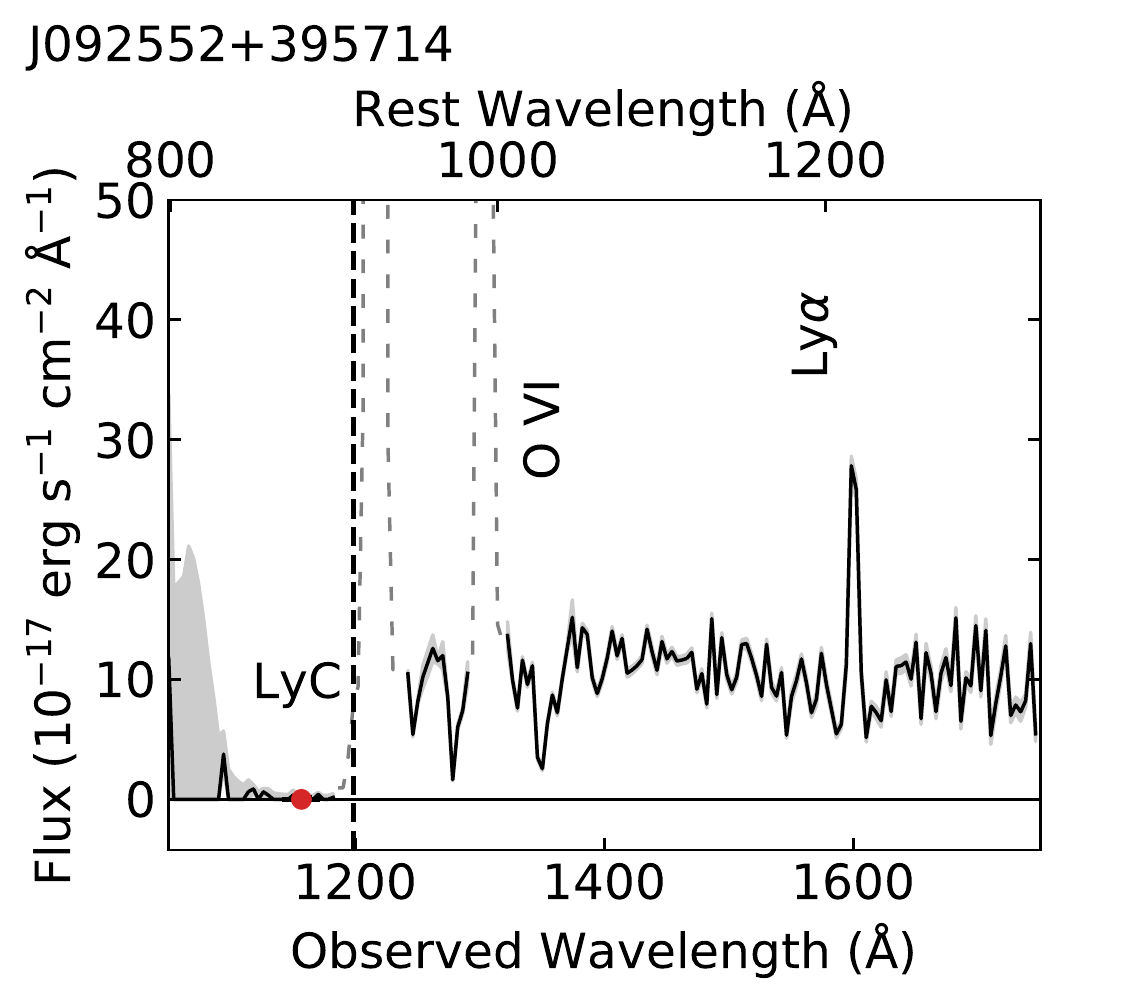}
    \includegraphics[width=0.32\textwidth]{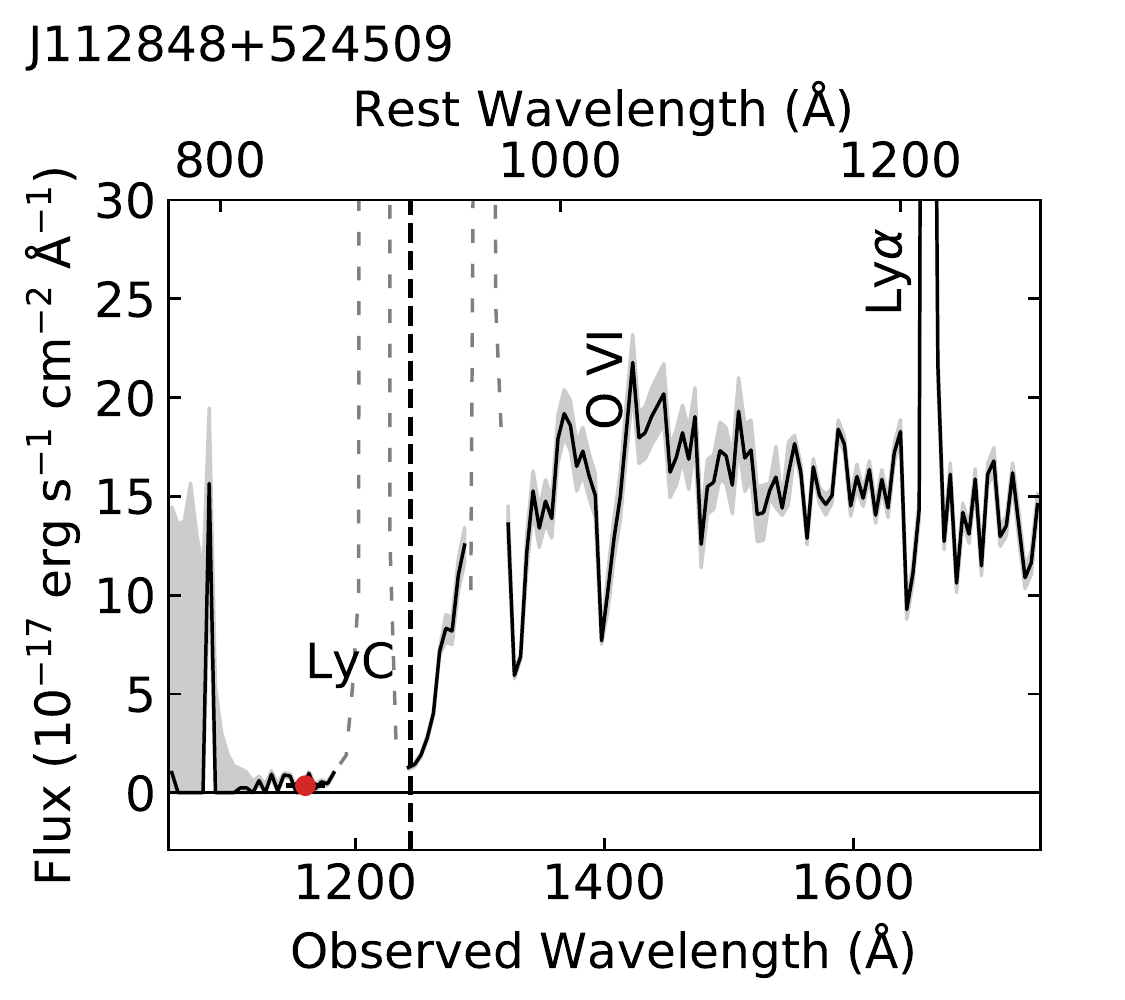}
    \includegraphics[width=0.32\textwidth]{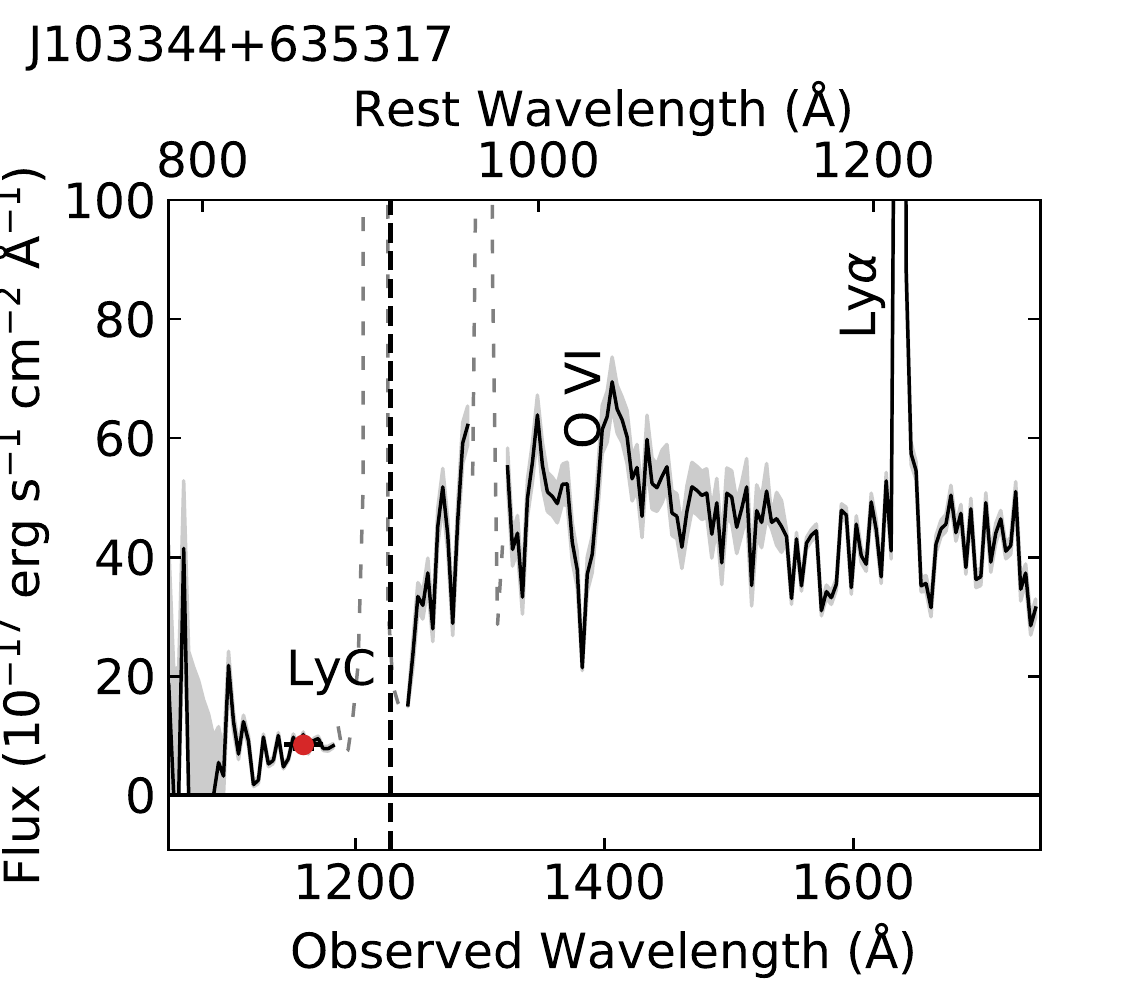}

    \includegraphics[width=0.32\textwidth]{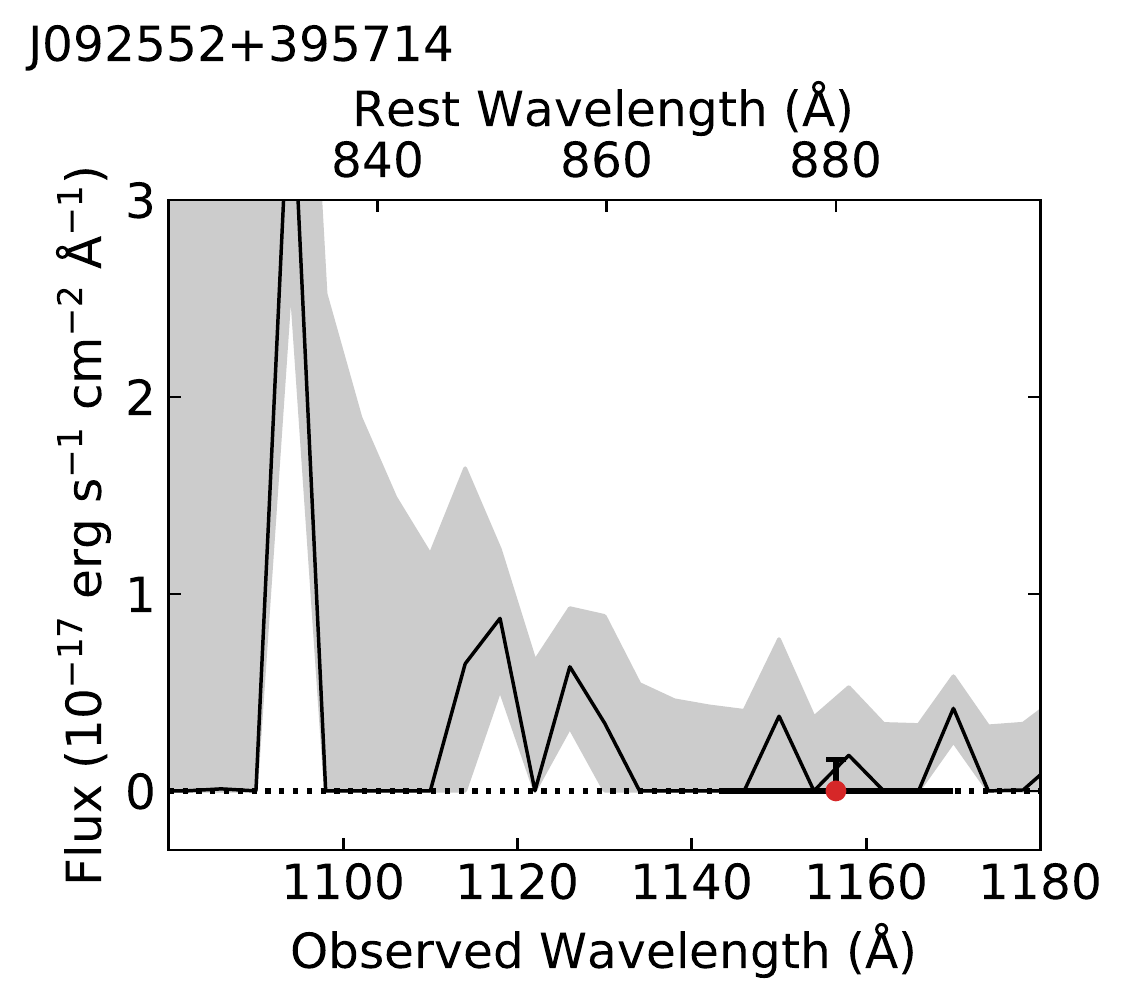}
    \includegraphics[width=0.32\textwidth]{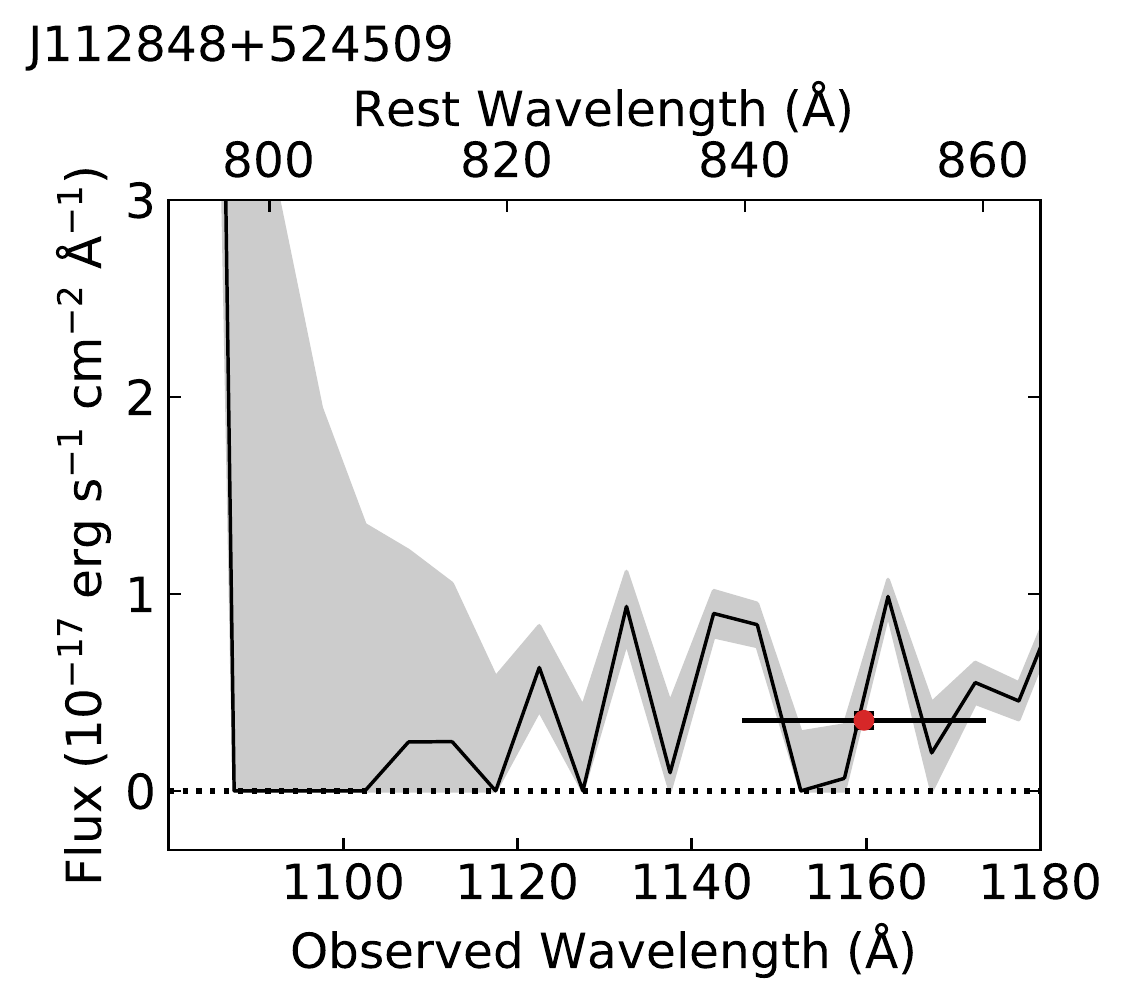}
    \includegraphics[width=0.32\textwidth]{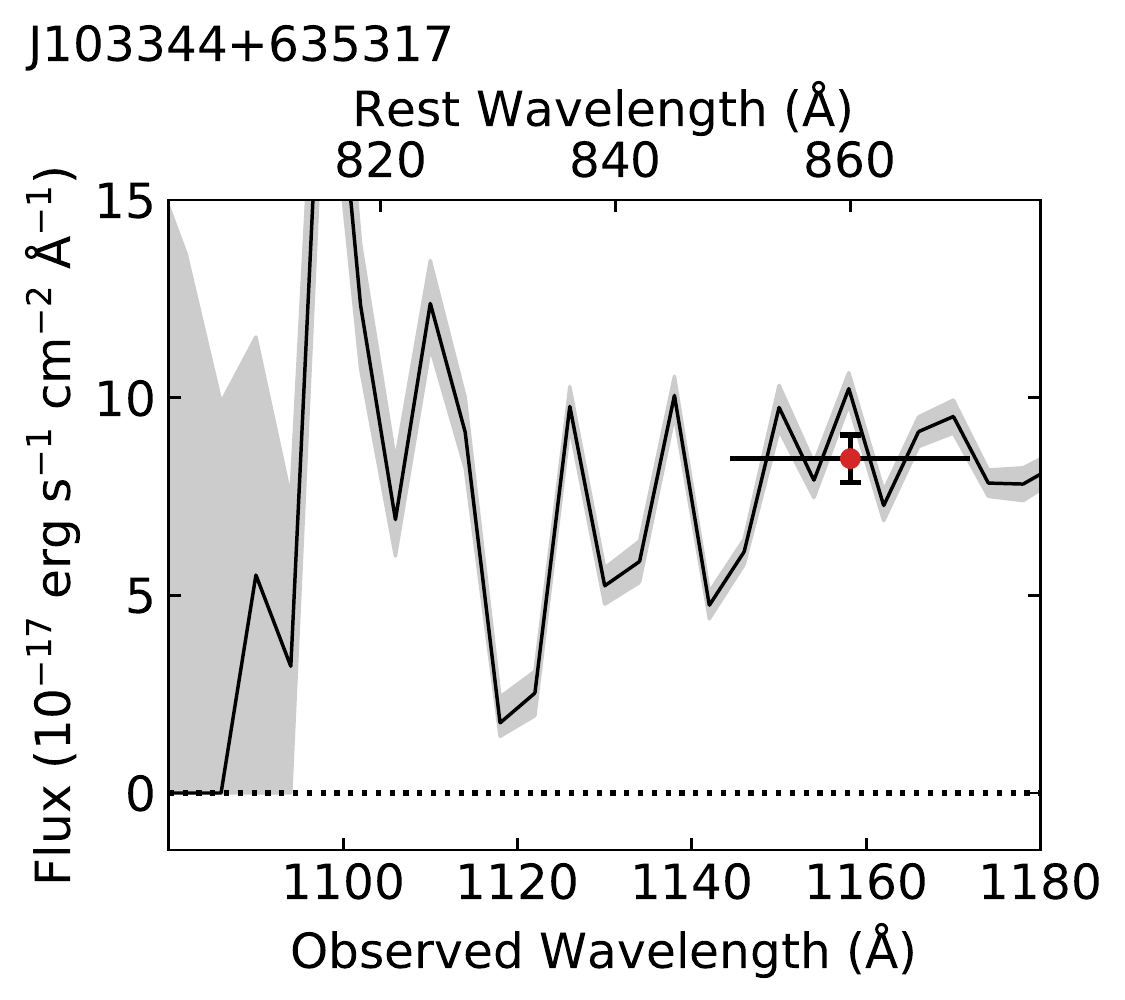}
    
\caption{{Top:} Example spectra (black) of a non-emitter ({left}), marginally detected Lyman continuum emitter (LCE, {center}) and well detected LCE ({right}) from the LzLCS, downsampled to 4 \AA\ resolution for visualization. Grey shaded region indicates the 1-$\sigma$ uncertainty in flux density. Red circle is the measured LyC flux with black lines showing the 68\% confidence intervals (capped) and rest-frame 20 \AA\ spectral bin (uncapped). Grey broken lines indicate the clipped geocoronal emission from telluric Ly$\alpha$ and \ion{O}{1}. Dashed vertical line indicates the Lyman limit. {Bottom:} Same as top but with LyC region magnified. Spectrum thumbnails for the full sample are available online.\label{fig:cos_spectra}}

\end{figure*}

\begin{figure*}
    \centering
    \includegraphics[width=0.32\textwidth]{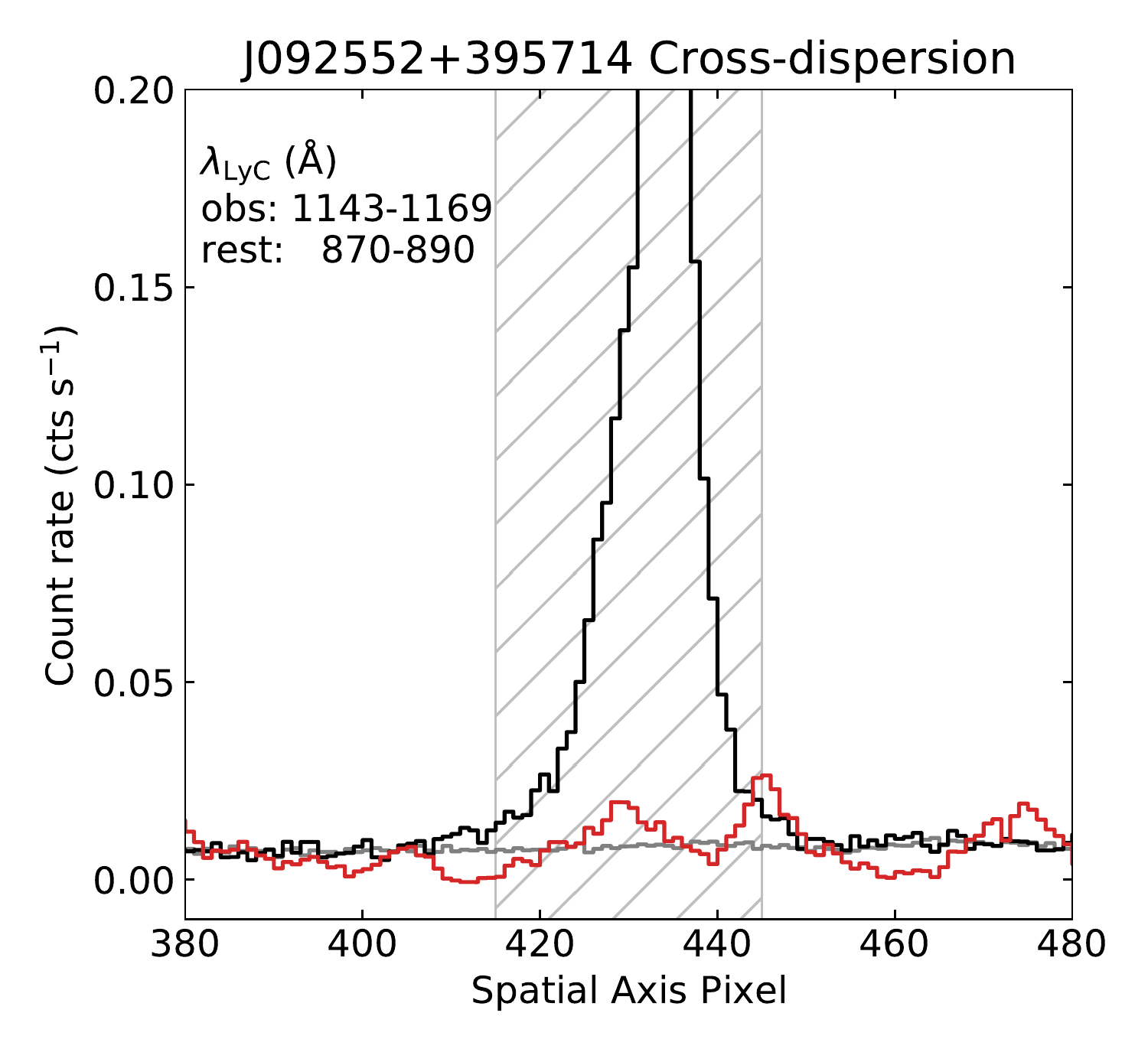}
    \includegraphics[width=0.32\textwidth]{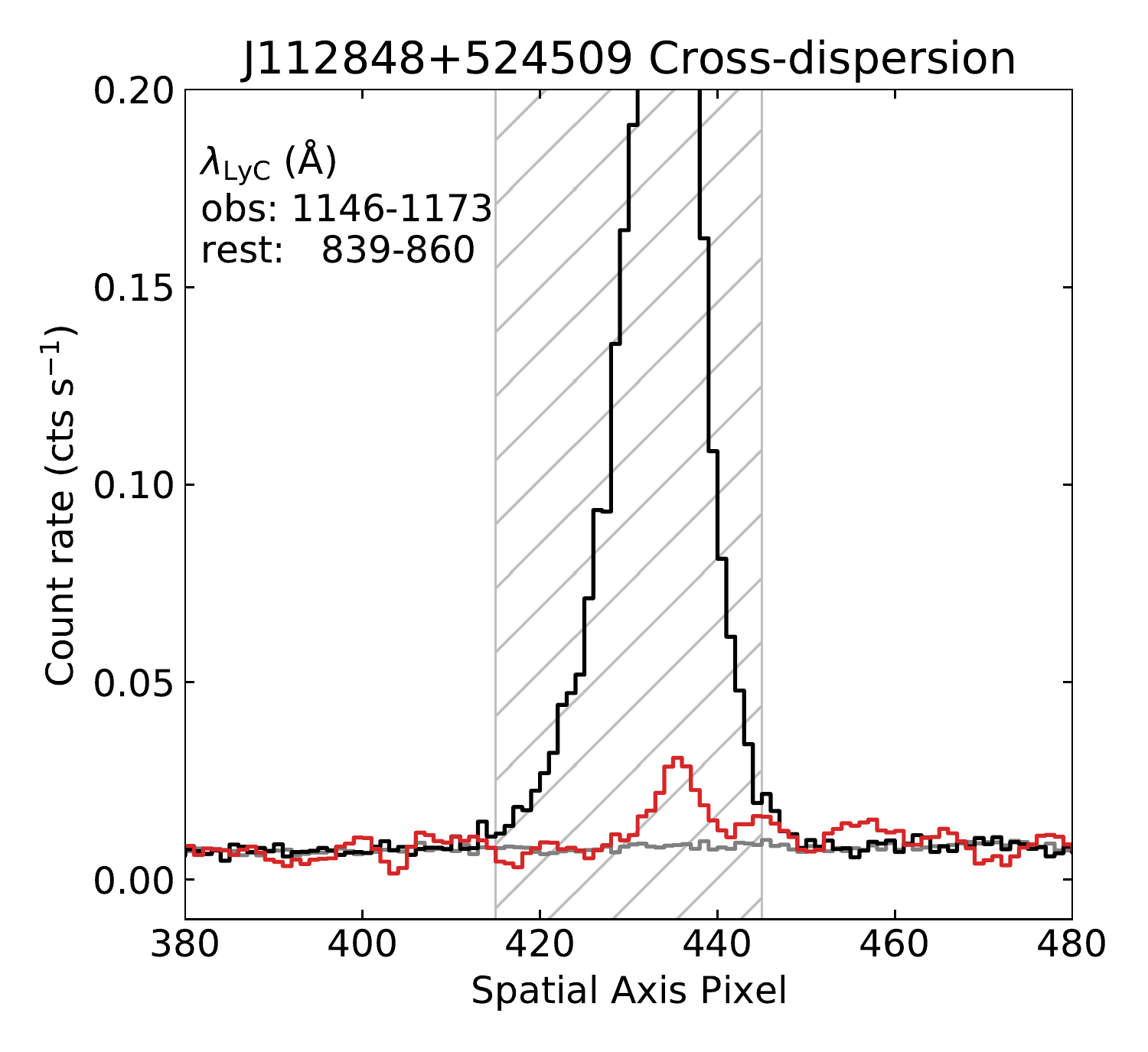}
    \includegraphics[width=0.32\textwidth]{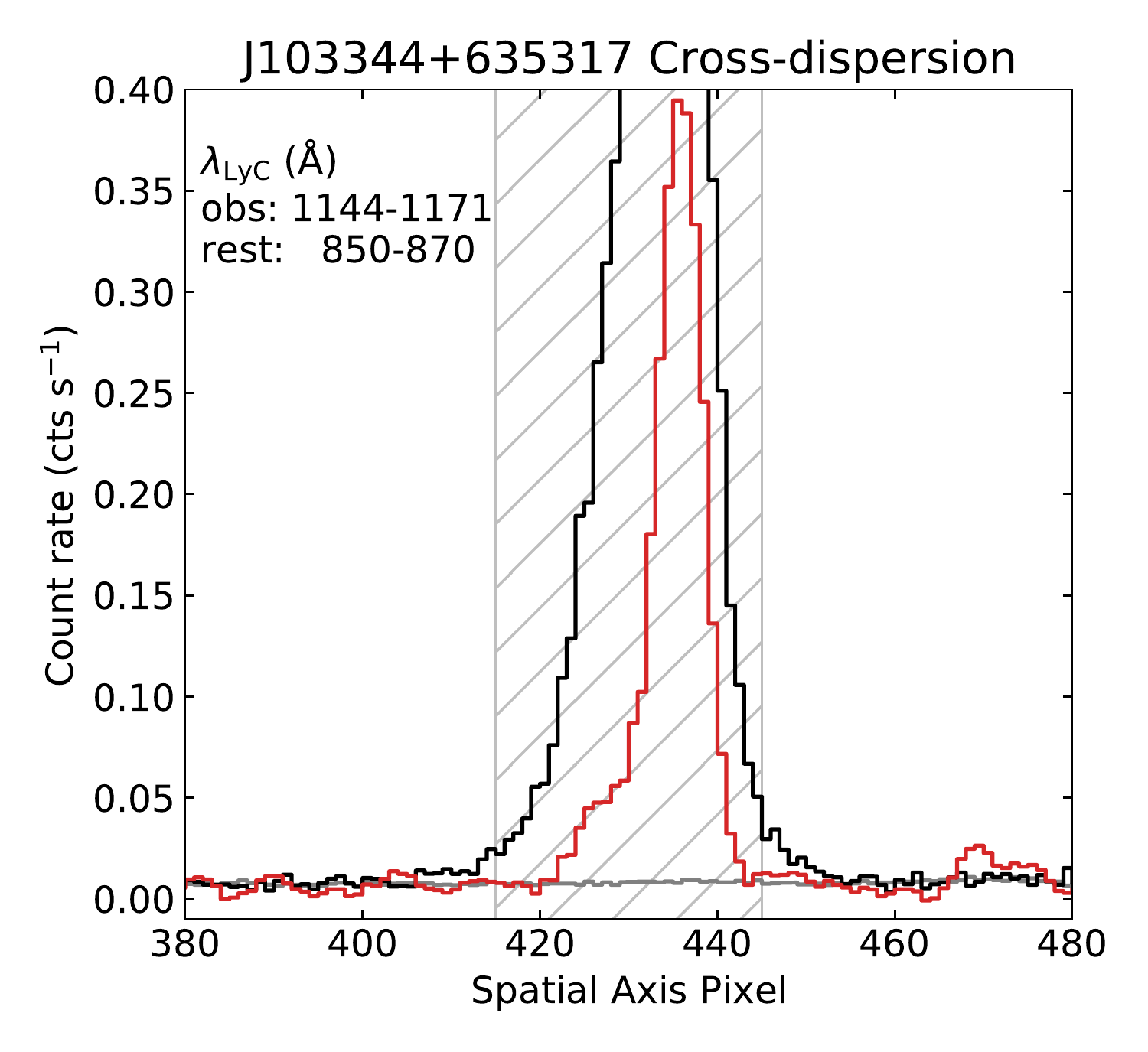}
    \caption{Example spatial cross-section of the 2d spectrum for a non-emitter ({left}), marginally detected Lyman continuum emitter (LCE, {center}) and well detected LCE ({right}) from the LzLCS. The black line represents the non-ionizing starlight continuum, the grey line the model dark current, red line the LyC scaled to the same median background counts as the starlight profile, and the grey hatched stripe the spectral extraction aperture.\label{fig:spec2d}}
\end{figure*}

We consider the mean background-subtracted flux density in the spectral window to be the LyC flux, $F_{\lambda \rm LyC}^{obs}$, and the 84th percentile in the background distribution (i.e., the $1\sigma$ sensitivity limit) to be the upper limit on $F_{\lambda \rm LyC}^{obs}$ in cases of non-detections. As a robust assessment of detection, we determine the probability, $P(>N|B)$, that the total observed, or gross, counts $N$ within the spectral window are realized from the distribution of background counts $B$. \citet{2016ApJ...825..144W} define $P(>N|B)$ as the survival function for the Poisson distribution of the background counts { evaluated at the gross observed counts. Proceeding with this convention, the probability that the gross measured LyC counts $N$ is a chance realization of the background $B$ is
\begin{equation}
    P(>N|B) = 1-\mathrm{Q}(N+1,B)
\end{equation}
where $\mathrm{Q}$ is the regularized incomplete gamma function.} We consider $P(>N|B)=0.02275$, the 2$\sigma$ value given by the normal distribution survival function, an acceptable maximum probability that the observed counts are sampled from the background.{ We list the number of galaxies in the LzLCS which satisfy traditional 2, 3, and 5$\sigma$ detection criteria in Table \ref{tab:detections} and provide examples of the rest-frame LyC for non-, weak ($\sim2\sigma$), and strong ($\gtrsim5\sigma$) detections in Figure \ref{fig:cos_spectra}.} In total, 35 galaxies satisfy our detection requirements.

\begin{figure}
    \centering
    \includegraphics[width=\columnwidth]{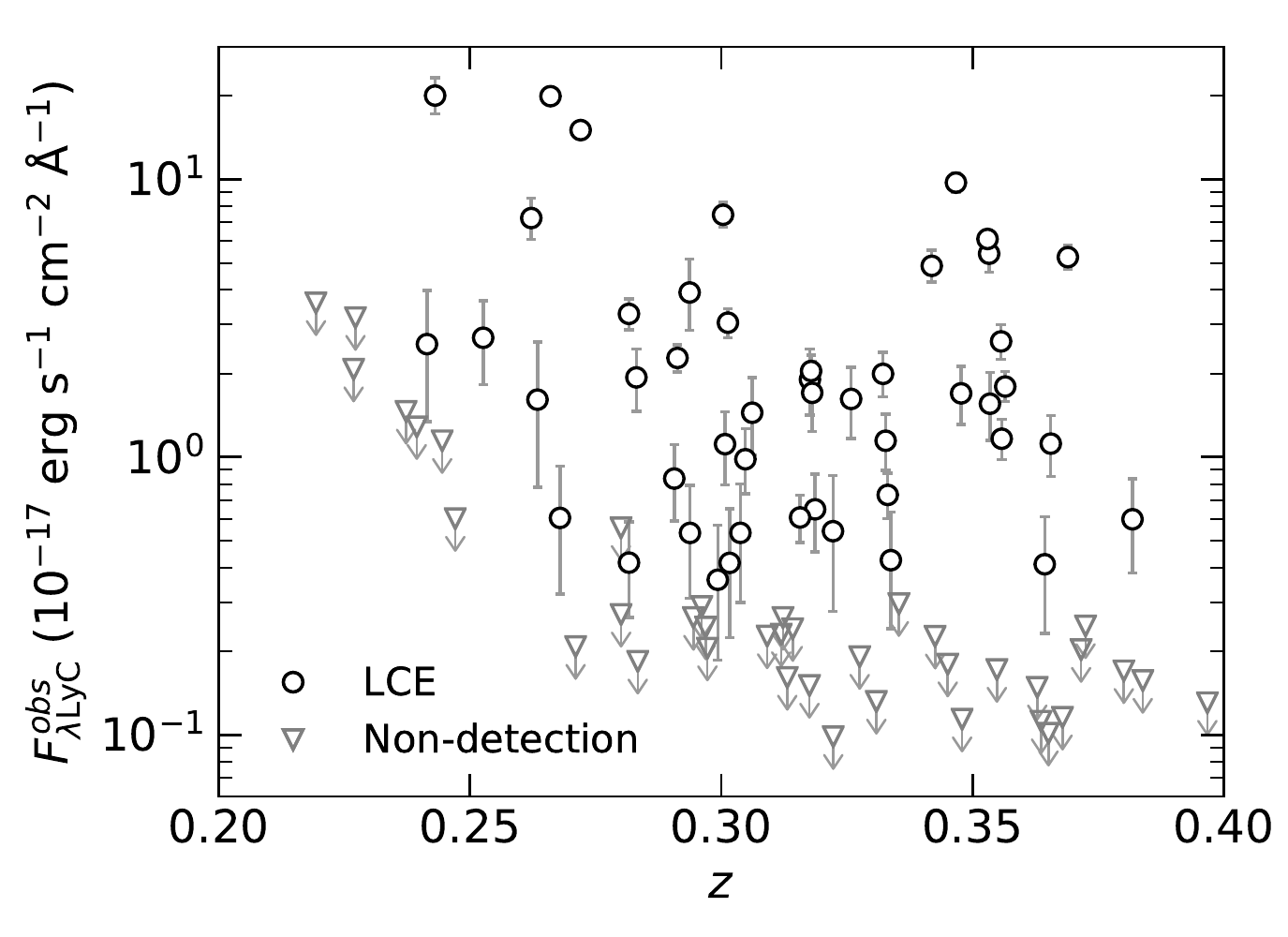}
    \caption{LyC measurements (circles) and upper limits (triangles) for the combined LzLCS and published samples show the {\it HST}/COS LyC flux sensitivity limits as a function of redshift. The increase in sensitivity with redshift is due {primarily to the location of the LyC on the COS detector.}\label{fig:cos_sensitivity}}
\end{figure}

To confirm these detections, we examine the cross-dispersion profile of the two-dimensional spectrum in the LyC window to qualitatively verify the presence of a LyC profile which appears roughly consistent with the non-LyC starlight profile. We show such a comparison of LyC with the non-LyC starlight profile in Figure \ref{fig:spec2d}. % Through this inspection, we find one marginal detection which may be produced by fluctuations in the dark current. 
In summary, we detect 35 LCEs out of 66 targets. { We show the LzLCS LyC fluxes in Figure \ref{fig:cos_sensitivity} as a function of redshift to highlight the effect of changes in COS sensitivity across the detector\footnote{ We estimate that pending changes to the COS flux calibration, which are not yet public for our Lifetime Position 4 settings, will result in a 6-6.5\% increase in our measured LyC flux and a smaller increase in flux at longer wavelengths. However, this increase is comparable to the relative uncertainty in the measured flux and will only increase the corresponding \fesclyc\ by a factor of $\sim1.02$.}.}%\citep[due to updates to the reference white dwarf models in][]{2020AJ....160...21B}

\begin{deluxetable}{lcccc}
\tablecaption{LzLCS and published detections of local LCEs with {\it HST}/COS. {Significance is the number of Poisson standard deviations between the measured LyC counts and the model background counts. $P(>N|B)$ is the probability of measuring LyC counts greater than those measured in the COS spectrum given the background counts. LzLCS indicates the number of galaxies from the LzLCS corresponding to each detection criterion. Pub. indicates the number of galaxies from {\it HST}/COS observations published in the literature and reprocessed by the LzLCS collaboration.} \label{tab:detections}}
\tablewidth{\textwidth}
\tablehead{
\colhead{Quality} & \colhead{Signif.} & \colhead{max $P(>N|B)$} & \colhead{LzLCS} & \colhead{Pub.}
}
\startdata
Good & $>5$ & $2.867\times10^{-7}$ & 12 & ~14\\
Fair & $3$-$5$ & $1.350\times10^{-3}$ & 13 & ~0\\
Marginal & $2$-$3$& $2.275\times10^{-2}$ & 10 & ~1\\
\hline
Detected & $>2$ & $2.275\times10^{-2}$ & 35 & 15 \\
Upper limit & $\le2$ & $1$ & 31 & 8\\
\enddata
\end{deluxetable}

To ensure consistency in our method, we reprocess the raw {\it HST}/COS spectra for the 23 galaxies in the \pubsamp\ investigations { of local ($z\lesssim0.4$) LCE candidates} following the same procedure and find that we reproduce their LyC fluxes. The median relative difference between the published fluxes and our measurements is $0.094_{-0.022}^{+0.043}$, indicating that we recover their results to within 10\% but with a statistically significant difference. While the scatter and differences are small, this slight disagreement suggests our reprocessing and re-measurement of the \pubsamp\ LyC fluxes are necessary to mitigate any systematic discrepancies. { Because we used the same custom reduction software, we attribute the difference to a combination of the spectral window over which we measure the LyC.} We find 15 of the 23 \pubsamp\ LCE candidates satisfy our detection requirements. { For the 23 LCE candidates, our significance assessment is consistent with the literature; however, we consider two objects, J124810+425954 \citep{2018MNRAS.478.4851I} and J112721+461042 \citep{2021MNRAS.503.1734I}, to be non-detections because our detection requirement is more stringent. Both we and the authors report $1\sigma$ LyC detection significance in these instances. We present the LyC fluxes for these published LCE candidates with the LzLCS results in Figure \ref{fig:cos_sensitivity}.}

{ Thus, the 35 detections in the LzLCS sample nearly triples the total number of confirmed LCEs in the local universe.}

\section{L\lowercase{z}LCS Galaxy Properties}\label{sec:global_props}

{Below, we detail the measurement and calculation of various properties of the LzLCS galaxies and { compare them to the same properties for the 23 local ($z\lesssim0.4$) galaxies discussed in \S\ref{sec:lyc_meas} with \emph{HST}/COS observations of the LyC published in \pubsamp.} Furthermore, we demonstrate the much broader range of the LzLCS properties relative to previous LCE surveys. We list these properties in Tables \ref{tab:optical_props}-\ref{tab:fesc} for several LzLCS targets and provide the full tables of these properties online.} 

\subsection{Optical Extinction}\label{sec:ebmv}

We have measured fluxes and equivalent widths (EWs) of emission lines in the SDSS DR15 spectra by fitting them with one to two Gaussian profiles following \citet{2019ApJ...885...96J}. In some cases, the [\ion{O}{3}]$\lambda$5007 profile appears ``sheared off'' or affected by sky lines, and the [\ion{O}{3}] $F_{5007}/F_{4959}$ flux ratio deviates significantly from the expected ratio of 2.98 \citep{2000MNRAS.312..813S}. In these cases, we adopt $F_{5007}=2.98F_{4959}$. We convert the observed EWs to the rest frame using redshifts obtained from the SDSS. Using the dust maps by \citet{2018MNRAS.478..651G} and the \citet{1999PASP..111...63F} extinction law, we correct the observed-frame emission line fluxes for Galactic extinction. Then, we iteratively compute the uncertainty-weighted rest-frame internal $E(B-V)$ and stellar absorption from the H$\alpha$, H$\beta$, H$\gamma$, H$\delta$, and H$\varepsilon$ emission line fluxes and EWs if the emission line flux is detected at S/N$>5$. Correction for stellar absorption is adapted from Equation 1 from \citet{1994ApJ...435..647I}. In the seven cases where the H$\alpha$/H$\beta$ is more than one standard deviation below the Case B value of 2.747 { \citep[][assuming the extreme case of $T_e=2\times10^4$ K and $n_e=10^2$ cm$^{-3}$]{1995MNRAS.272...41S}} and the four cases where the H$\alpha$ profile appears sheared off (likely due to spurious cosmic ray clipping or saturation), we exclude H$\alpha$ from the procedure.

We iterate the following procedure until converging on a solution for $E(B-V)$: (i) derive the electron temperature and density from nebular lines, (ii) calculate the intrinsic flux ratios from the results of step (i) by interpolating over the grid of recombination coefficients from \citet{1995MNRAS.272...41S}, and (iii) compute the variance-weighted average $E(B-V)$ from the ratio of observed to intrinsic Balmer decrements. Electron temperature and density derivation follows the temperature-scaling approach described by \citet{2020MNRAS.496.2191F} with collisional populations and emissivities computed by {\sc pyneb} \citep{2015A&A...573A..42L}. For the electron temperature, we use the [\ion{O}{3}] $\lambda\lambda4363;4959,5007$\ auroral line and nebular doublet, substituting the [\ion{O}{3}] $\lambda4363~$ flux inferred from the ``ff--relation'' by \citet{2006MNRAS.367.1139P} for the 12 galaxies in the total LzLCS sample where the auroral line is not detected \citep[see discussion in][]{2017MNRAS.465.1384C}. For the electron density, we use the [\ion{S}{2}] $\lambda\lambda6716,31$\ doublet, available for 56 galaxies, as the [\ion{O}{2}] $\lambda\lambda3726,29$\ doublet is not resolved in the SDSS spectra. Otherwise, we assume $n_e=100$ cm$^{-3}$. % as this is characteristic of GP galaxies \citep[e.g.][]{2013ApJ...766...91J}.%z We explore the effects of varying the internal reddening curve by repeating this iterative process for three different extinction laws: Galactic-like \citep{1989ApJ...345..245C}, starburst-like \citep{2001PASP..113.1449C}, and SMC-like \citep{2003ApJ...594..279G}. Since the reddening law is not well constrained for our sample, we proceed with the fluxes corrected using the \citet{1989ApJ...345..245C} extinction curve as this choice gives the median $E(B-V)$ values and appears appropriate for the GPs \citep{2017MNRAS.467.4118I}. While the SMC-like extinction is comparable to the Galactic-like extinction ($\Delta E(B-V)=0.0004\pm0.006$), the starburst-like extinction differs by $\Delta E(B-V)=-0.014\pm0.020$.
We assume the \citet{1989ApJ...345..245C} extinction law as this choice gives $E(B-V)$ values comparable to other extinction laws \citep[e.g.,][]{2003ApJ...594..279G} and appropriately describes the extinction of nebular emission lines in LCEs such as the GPs \citep{2017MNRAS.467.4118I}.

\begin{figure}
    \centering
    \includegraphics[width=\columnwidth]{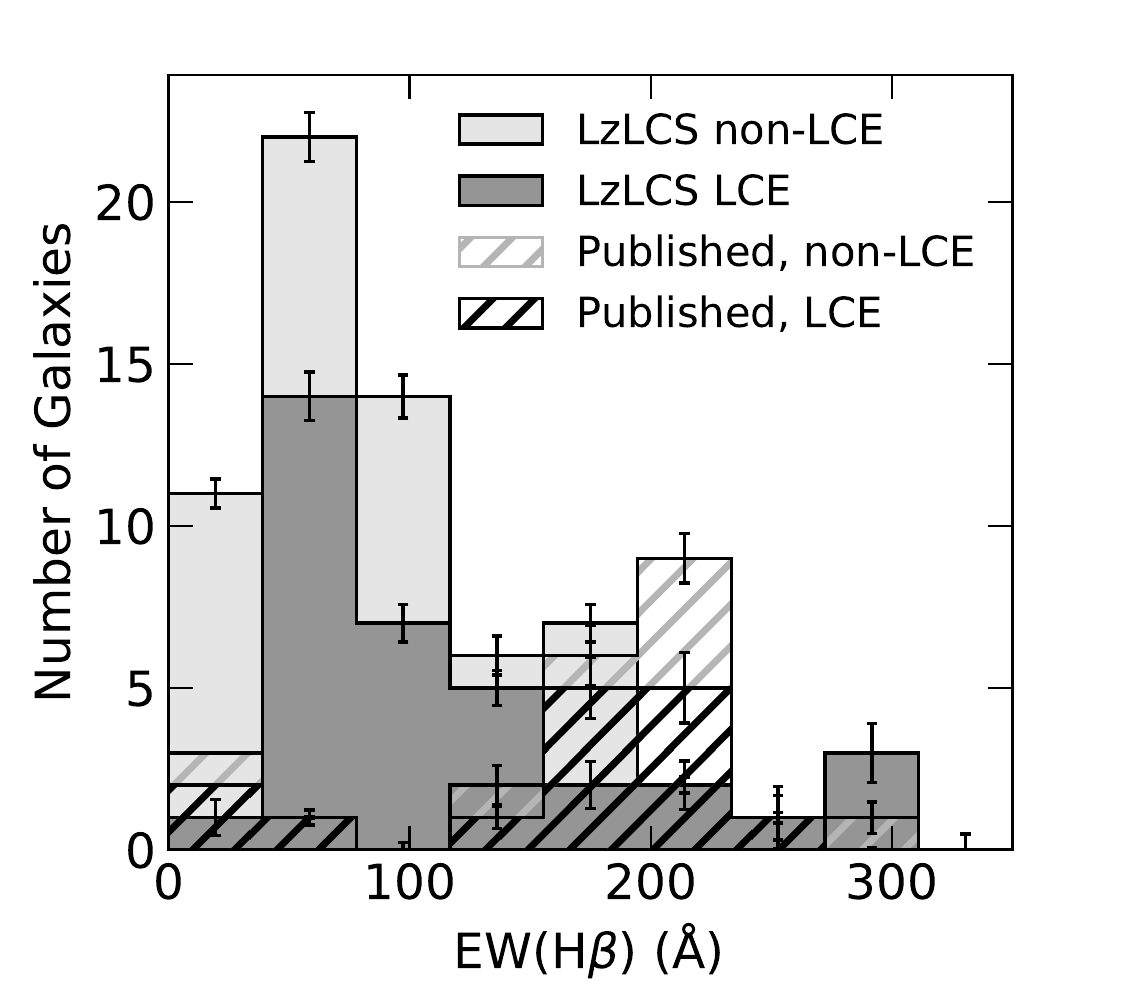}
    \caption{Distribution of rest-frame H$\beta$ equivalent width (EW) values for galaxies in the LzLCS (solid) with detected LyC (dark grey) and undetected LyC (light grey). For comparison, we include the EWs for LCE candidates with published \emph{HST}/COS spectra (hatched) with detected LyC (black) and undetected LyC (grey). Error bars represent the $1\sigma$ Poisson binomial uncertainty in each histogram bin.\label{fig:ewhb_hist}}
\end{figure}

\begin{figure}
    \centering
    \includegraphics[width=\columnwidth]{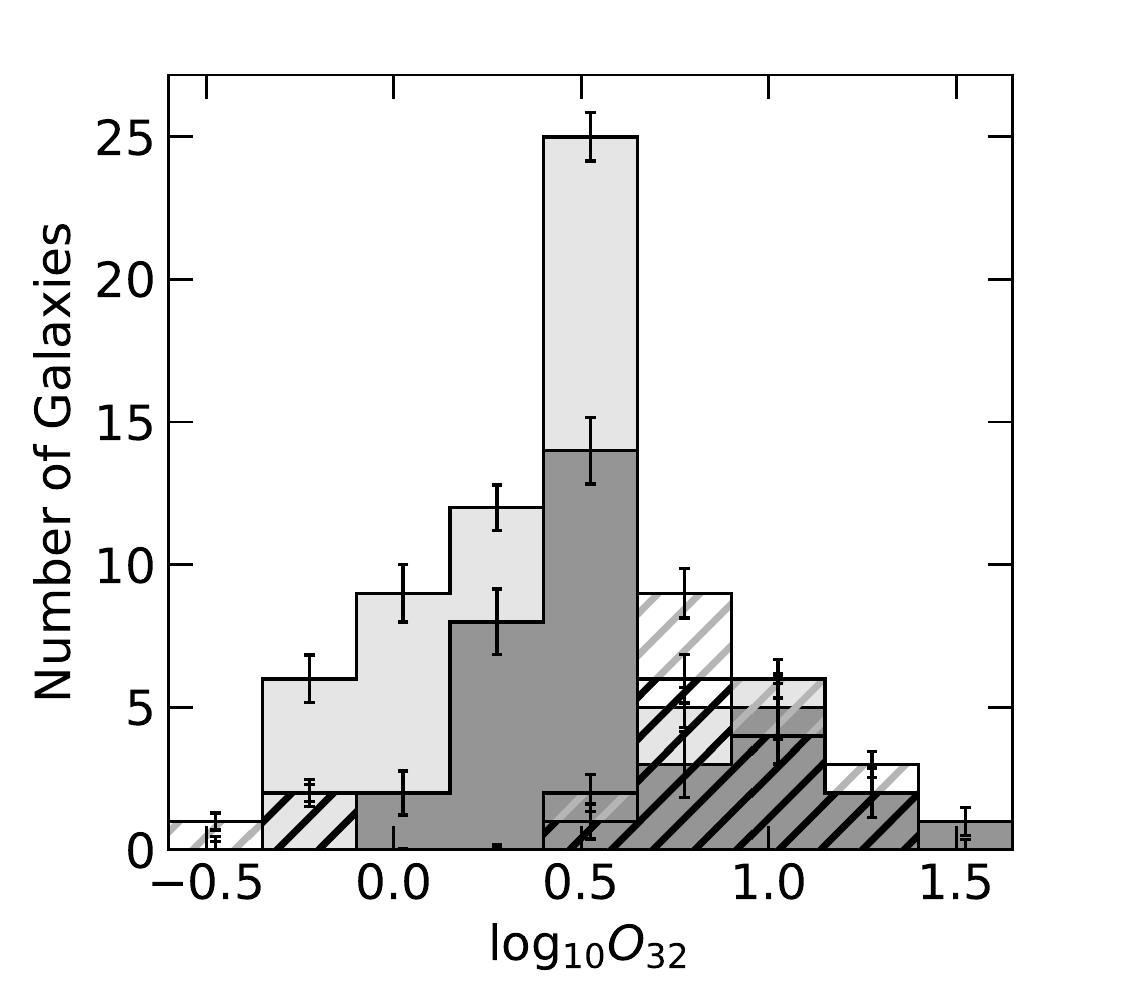}
    \caption{Same as Figure \ref{fig:ewhb_hist} but for [\ion{O}{3}]$\lambda5007$/[\ion{O}{2}]$\lambda3727,29$=\orat.\label{fig:o32_hist}}
\end{figure}

{These corrected flux measurements from the optical spectra provide the H$\beta$ EW (accounting for stellar absorption) and \orat\ flux ratios (accounting for extinction).} We show the H$\beta$ EW for the LzLCS galaxies and published LCEs in Figure \ref{fig:ewhb_hist}. Values range from 11 to 426 \AA\ with a median of 91 \AA. For the LzLCS galaxies and published LCEs, $\log_{10}$\orat\ spans -0.32 to 1.56 with a median of 0.65, as we show in Figure \ref{fig:o32_hist}. From both figures, the previously published LCE H$\beta$ EWs and \orat\ ratios are located towards the high end of the LzLCS sample distribution. In other words, the LzLCS extends to much lower H$\beta$ EWs and \orat\ than these previous studies. { The LzLCS nearly doubles the number of detected low-redshift LCEs with H$\beta$ EWs $>100$ \AA\ and \orat\ $>3$. Moreover, the LzLCS dramatically improves the number of detected low-redshift LCEs with H$\beta$ EWs $<100$ \AA\ and \orat\ $<3$, a space previously sampled by just 3 LCEs. The presence of LCEs across such a wide range of EW H$\beta$ suggests galaxies with a variety of burst ages and/or star formation histories can leak LyC photons. Similarly, the presence of LCEs across nearly 2 dex in \orat\ indicates that LCEs span a wide range of ionization parameters and/or nebula boundary conditions.}

\subsection{Nebular Abundances}\label{sec:nebabn}

\begin{figure}
    \centering
    \includegraphics[width=\columnwidth]{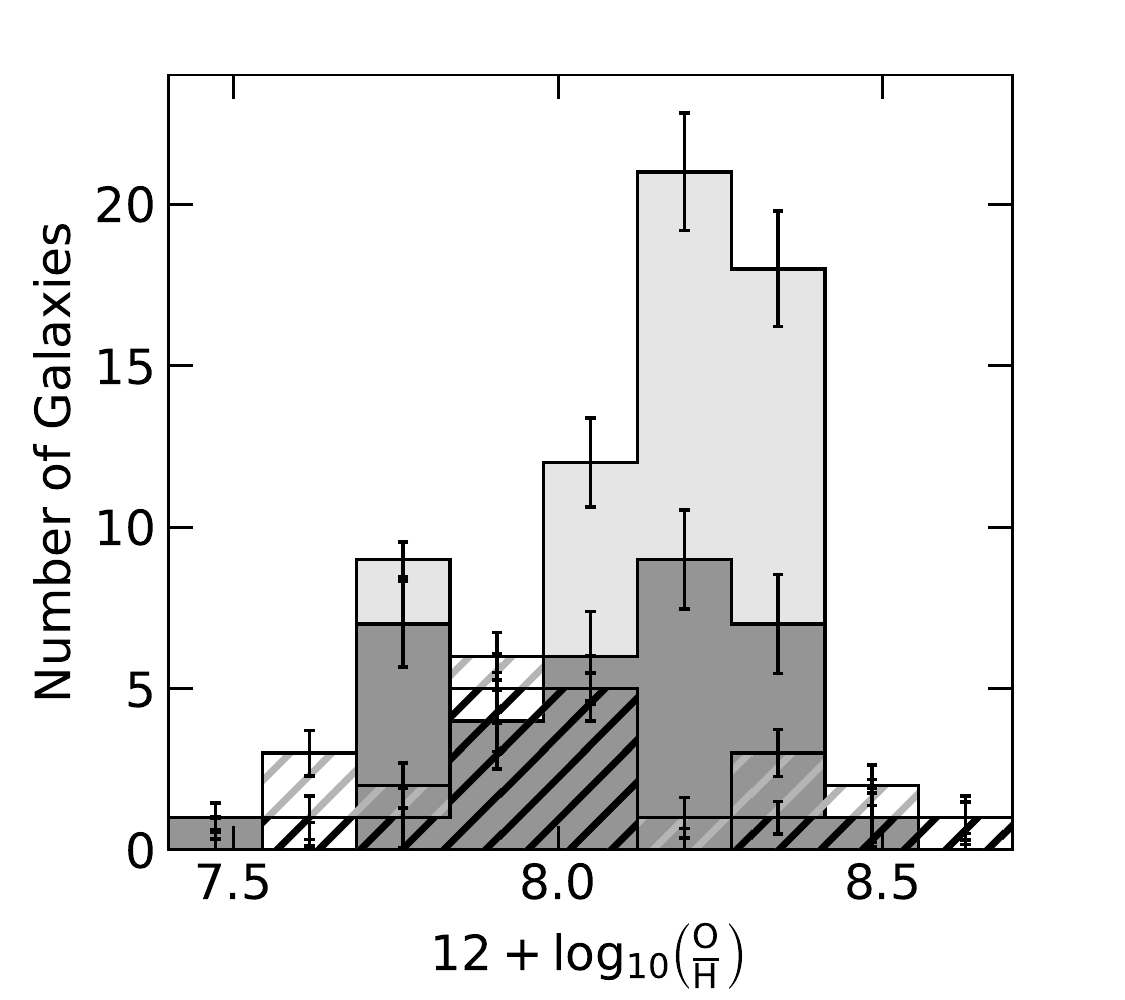}
    \caption{Same as Figure \ref{fig:ewhb_hist} but for $12+\log_{10}({\rm O/H})$.\label{fig:oh12_hist}}
\end{figure}

With the electron temperatures and densities derived above, we determine direct-method relative oxygen abundances from emissivities computed by {\sc pyneb} and extinction-corrected optical emission lines. Monte Carlo sampling the emissivities using the uncertainties in the fluxes, temperatures, and densities yields the total statistical uncertainty in our direct-method abundances. As evident in Figure \ref{fig:oh12_hist}, the LzLCS spans a range of about 6 to 60\% solar oxygen abundance \citep[as defined by][]{2015A&A...583A..57S} while the published LCEs are more narrowly concentrated to a range of 10 to 30\% solar. The LzLCS samples higher metallicities than previous studies, with $\sim$50\% of the galaxies having $12+\log_{10}({\rm O/H })$ above the highest published local LCEs. { Furthermore, the LzLCS increases the number of detected low-redshift LCEs across all abundances, particularly above $12+\log_{10}({\rm O/H })\sim8.2$, demonstrating that, as with H$\beta$ EWs, LCEs span a wide range of of star formation histories because $12+\log_{10}({\rm O/H })$ traces the net number of type II SNe.}

\subsection{Half-light Radius}\label{sec:halfrad}

Using the reduced MIRROR-A NUV COS acquisition images, we compute the source radius $r_{50}$ containing 50\% of the background-subtracted counts. We estimate the median background counts in an annulus centered on the source with an inner radius of 53 pixels to avoid contamination by the source and subtract the median counts from the image. {After background subtraction, we correct for vignetting effects. While the galaxies in the LzLCS are typically compact (typical uncorrected galaxy profile FWHM$\lesssim0.4$\arcsec\ in the NUV), we correct the acquisition images for the radial decline in throughput because the total source counts can still be affected by vignetting.} Then, we compute the total source counts by measuring the counts enclosed by a range of radii until the total counts vary by less than the rms of the background noise. We then interpolate over the counts distribution to obtain the radius at half the total source counts. Uncertainty in the half-light radius is determined by summing the Poisson error of the gross counts, the COS acquisition image plate scale of 0.0235\arcsec\ px$^{-1}$, and a maximum NUV imaging PSF FWHM of 2.4 px.

\begin{figure}
    \centering
    \includegraphics[width=\columnwidth]{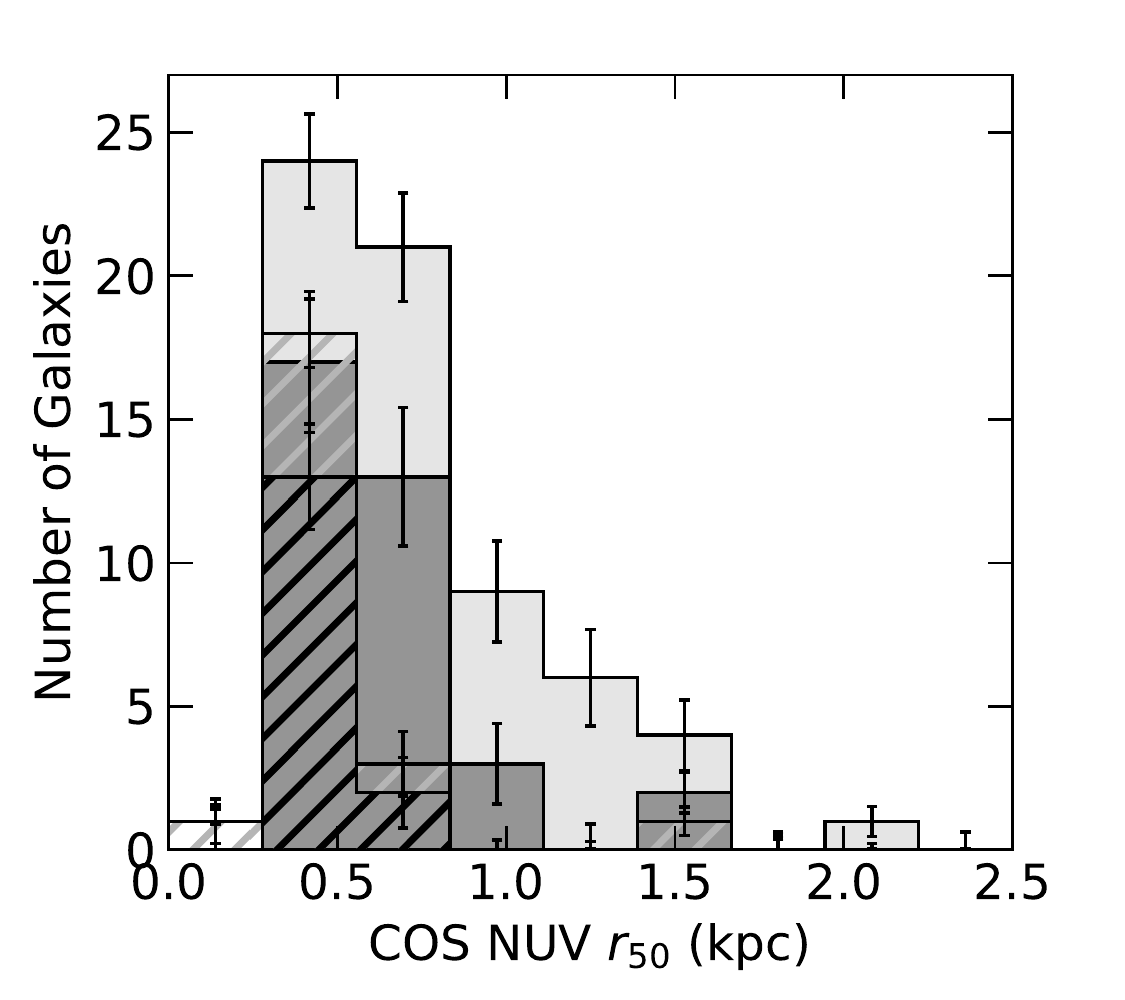}
    \caption{Same as Figure \ref{fig:ewhb_hist} but for $r_{50}$.\label{fig:r50_hist}}
\end{figure}

In Figure \ref{fig:r50_hist}, we show that the UV-emitting stellar populations inhabit small regions with $0.3<r_{50}<0.6$ kpc in both the published LCEs and half of the LzLCS galaxies. While the number of galaxies in each bin decreases quickly with increasing halflight radius, the LzLCS galaxies have $r_{50}$ as high as $2.25$ kpc, indicating the survey includes galaxies with spatially-extended star formation. { However, as in previous studies, the LzLCS finds that LCEs predominantly have { compact star-forming regions}. To confirm this result, we} also fit the surface brightness distributions with two-dimensional S\'ersic and exponential profiles and find close agreement between the best-fit effective radii and the model-independent halflight radii.

\subsection{Ly$\alpha$ and Continuum Properties}\label{sec:uvprops}

The {\it HST}/COS spectra are extracted and reduced a second time following the procedure discussed in \S\ref{sec:obs}, { this time using a 30-pixel aperture (0.637\arcsec) in place of the 25-pixel aperture to include more signal because Ly$\alpha$ is more spatially extended than the UV continuum \citep[e.g.,][]{2015A&A...576A..51G,2016A&A...587A..98W,2017A&A...608A...8L,2021arXiv211001626R} and requires less background exclusion than the LyC. For the LzLCS, the Ly$\alpha$ extraction aperture radius corresponds to a factor of about 2.6 more than the UV continuum halflight radius and thus should contain most of the Ly$\alpha$ flux \citep[e.g.,][]{2013ApJ...765L..27H}, although vignetting of the COS aperture may exclude some of the Ly$\alpha$ even in the more compact sources.} We process the data using the same assumptions as in the previous section, most notably the same Galactic extinction. From these wider extractions, we measure the integrated galactic Ly$\alpha$ flux. 
We fit the continuum within 100 \AA\ of Ly$\alpha$ with a linear fit using iterative sigma clipping {to exclude noise spikes and absorption features}, conservatively assuming a 25\% uncertainty in the continuum fit. We then integrate the continuum-subtracted flux density where the Ly$\alpha$ feature deviates from the continuum to obtain the Ly$\alpha$ flux, masking the 1206 and 1240 \AA\ features to avoid contamination. To obtain the rest-frame Ly$\alpha$ EW, we divide by the continuum flux and correct for redshift. {We do not correct for stellar Ly$\alpha$ absorption as its effect on the measured emission line flux and EW is, at most, relatively small \citep{2013AJ....146..158P}.}

\begin{figure}
    \centering
    \includegraphics[width=\columnwidth]{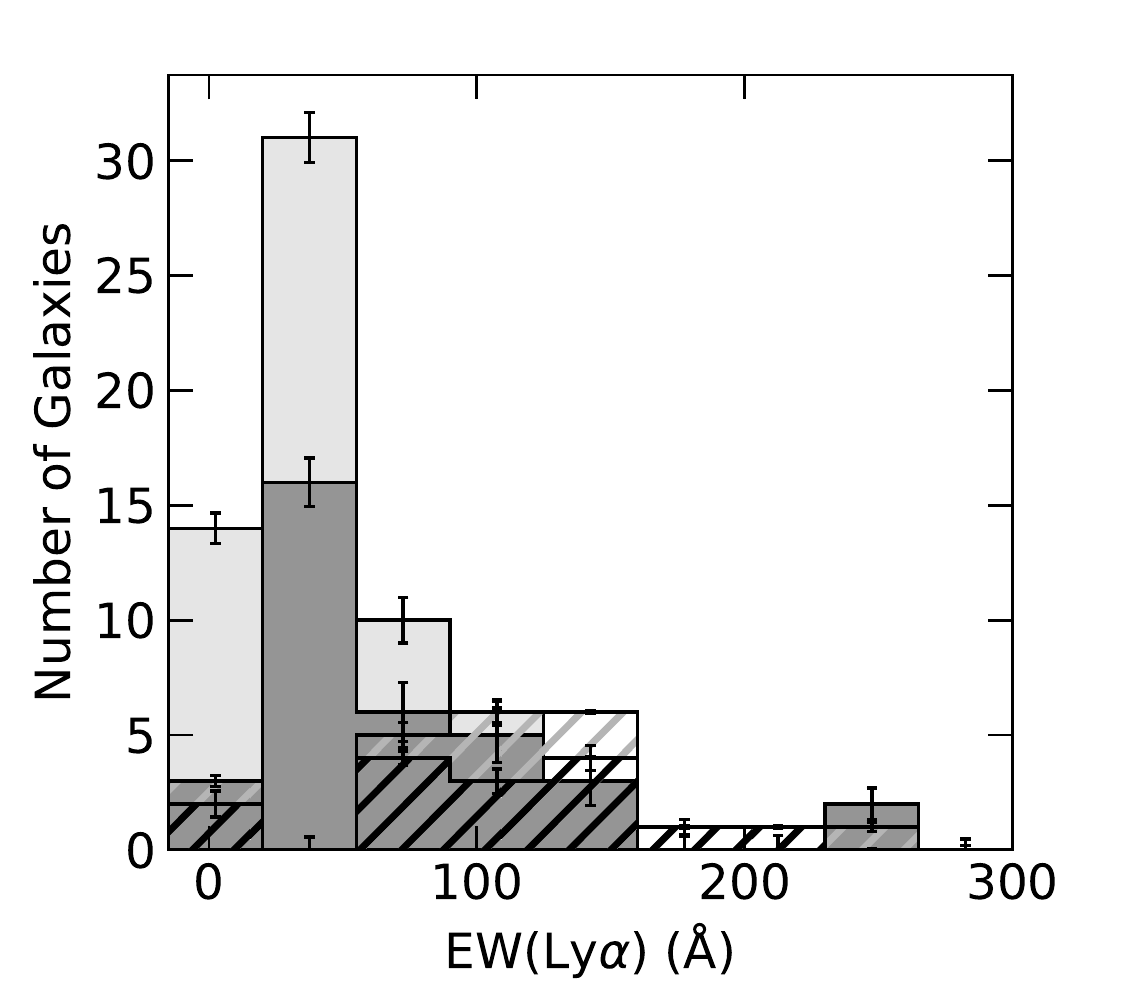}
    \caption{Same as Figure \ref{fig:ewhb_hist} but for rest-frame Ly$\alpha$ EW.\label{fig:ewlya_hist}}
\end{figure}

In Figure \ref{fig:ewlya_hist}, we show the Ly$\alpha$ EWs for the LzLCS and \pubsamp\ LCE candidates. The LzLCS Ly$\alpha$ EWs $\gtrsim60$ \AA\ are consistent with the majority of published LCEs; however, 45 of the 66 galaxies in the LzLCS sample have EWs smaller than this value. { As with H$\beta$, we find that LCEs span a wider range in Ly$\alpha$ EWs than previous studies. Because Ly$\alpha$ is more sensitive to the \ion{H}{1} column density and the corresponding continuum is more sensitive to recent star formation, the LzLCS results demonstrate that LCEs span a wider range of burst ages and/or \ion{H}{1} opacities than previously published surveys may have indicated.} 

\begin{figure}
    \centering
    \includegraphics[width=\columnwidth]{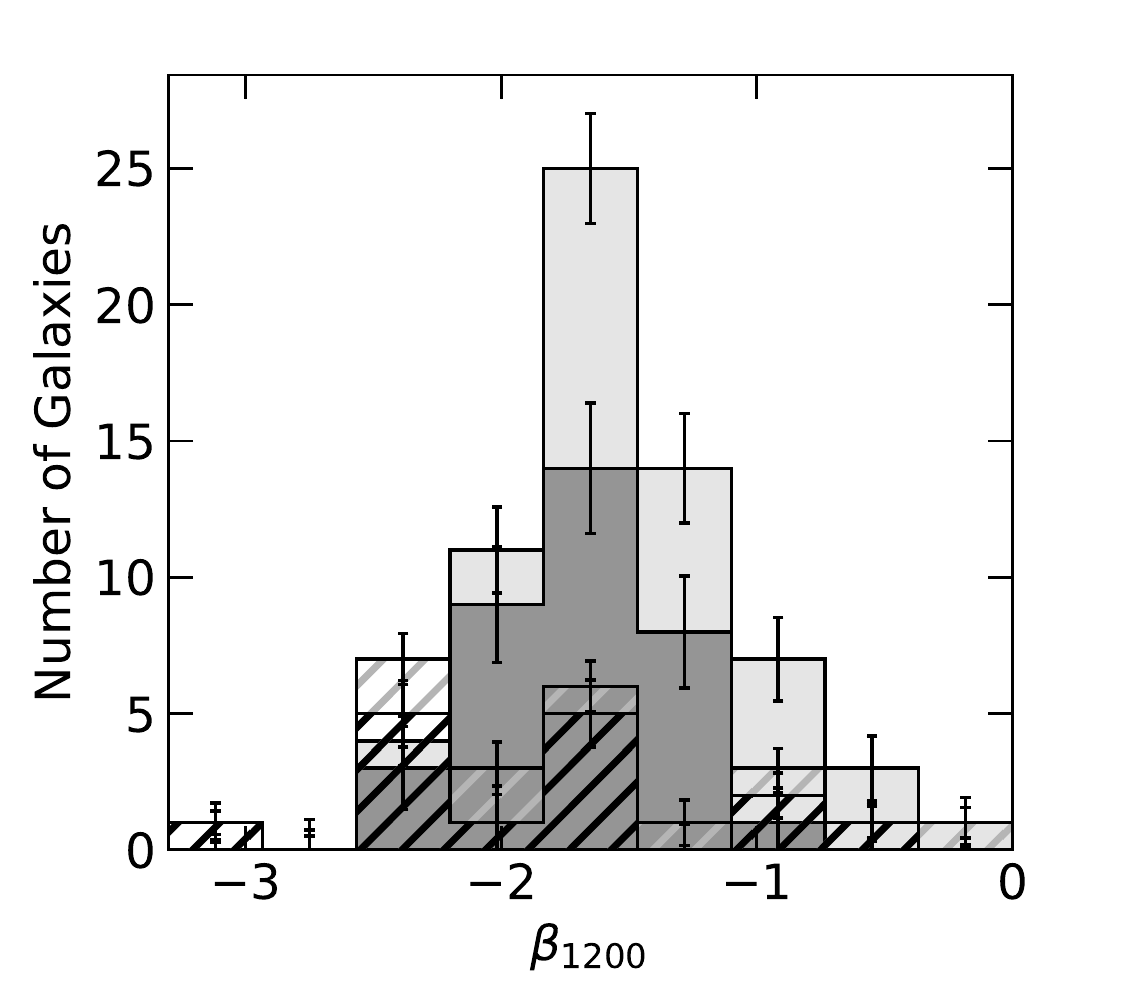}
    \caption{Same as Figure \ref{fig:ewhb_hist} but for $\beta_{1200}$.\label{fig:uvbeta_hist}}
\end{figure}

To determine the spectral index $\beta_{1200}$ of the rest-frame UV continuum, we fit the wide-extraction spectra using the affine-invariant Markov chain Monte Carlo sampling software {\sc emcee} \citep{2013PASP..125..306F} to sample the posterior of $\beta_{1200}$. We set a lower limit of 1050 \AA\ in the rest frame to avoid the combined \ion{O}{6} $\lambda\lambda1032,1038$\ \AA\ and \ion{C}{2} $\lambda1036$\ \AA\ absorption features and mask the galactic Ly$\alpha$. We show the distribution of $\beta_{1200}$ values in Figure \ref{fig:uvbeta_hist}. The LzLCS samples roughly the same range of $-2.5<\beta_{1200}<-0.35$ as previous studies of local LCEs but with most galaxies concentrated at $\beta_{1200}=-1.6$. One exception, J131904+510309, exhibits substantial extinction in the UV---$E(B-V)_{UV}$=0.5, $E(B-V)_{neb}$=0.42---which results in $\beta_{1200}=0.79$. { While the LzLCS LCEs span a range in UV $\beta$ comparable to that of previously published LCE candidates, our LCEs are, on average, redder, suggesting that LCEs can have a larger range of burst ages and/or dust content than previously found.}

%From the narrow-extraction spectra, we measure the integrated EWs of \ion{Si}{2} $\lambda1260$\ \AA\ and \ion{C}{2} $\lambda1334$ using a spline fit to the continuum, $c_\lambda$, and summing $1-f_\lambda/c_\lambda$ as in \citet{2017A&A...605A..67C}. We show these absorption line EWs in Figure \ref{fig:c2si2_hist}.% From Figure \ref{fig:c2si2_hist}, these absorption features are consistently optically thick when detected ({ I found this using the oscillator strengths and the curve of growth, but this might be extraneous?}).%, consistent with line-of-sight column densities on the order of $N\sim10^{16}-10^{17}$ cm$^{-2}$ from curve of growth estimates.

\iffalse
\begin{figure}
    \centering
    \includegraphics[width=\columnwidth]{figures/histograms/ew_c2_hist-LyC.pdf}
    \includegraphics[width=\columnwidth]{figures/histograms/ew_si2_hist-LyC.pdf}
    \caption{Same as Figure \ref{fig:ewhb_hist} but for \ion{C}{2} (top) and \ion{Si}{2} EWs (bottom).\label{fig:c2si2_hist}}
\end{figure}
\fi

Because the flux at 1500 \AA\ often falls outside the COS window for these redshifts, we also measure the continuum flux at 1100 \AA\ as other studies have done \citep[e.g.,][]{2019ApJ...885...57W}. This spectral window is a reliably bright part of the starlight continuum which avoids the aforementioned absorption features spanning 1030-1040 \AA. We take $F(1100)$ to be the average flux from 1090 to 1110 \AA, the same width as the LyC flux for consistency. At our sample's redshifts, this choice also serves to eliminate potential contamination of $F(1100)$ by telluric \ion{O}{1} $\lambda1304$ emission.

\subsection{UV Spectral Modeling}

Following \citet{2019ApJ...882..182C}, we estimate stellar $E(B-V)$ values and \fesclyc\ by comparing our data with a library of synthetic spectra compiled from {\sc Starburst99} models \citep{2010ApJS..189..309L} for non-rotating stars and nebular continuum modeled by {\sc Cloudy} \citep{2013RMxAA..49..137F}. These models are fit to the continua with scaling factors multiplied by the \citet{2016ApJ...828..107R} extinction law with reddening as an additional free parameter. Our reference library contains forty synthetic spectra for simple stellar populations spanning ten burst ages (1, 2, 3, 4, 5, 8, 10, 15, 20 and 40 Myr) and four metallicities (0.05, 0.2, 0.4 and 1 $Z_\odot$). These are combined allowing for multiple generations of star formation with metallicity as a free parameter (Salda\~na-Lopez et al. submitted). Fits are performed in the rest-frame after convolving model spectra by a Gaussian kernel to the COS spectral resolution. We discuss the derivation of \fesclyc\ further in \S\ref{sec:fesc}.

\subsection{Star Formation Rate Surface Density and Stellar Mass}

\begin{figure}
    \centering
    \includegraphics[width=\columnwidth]{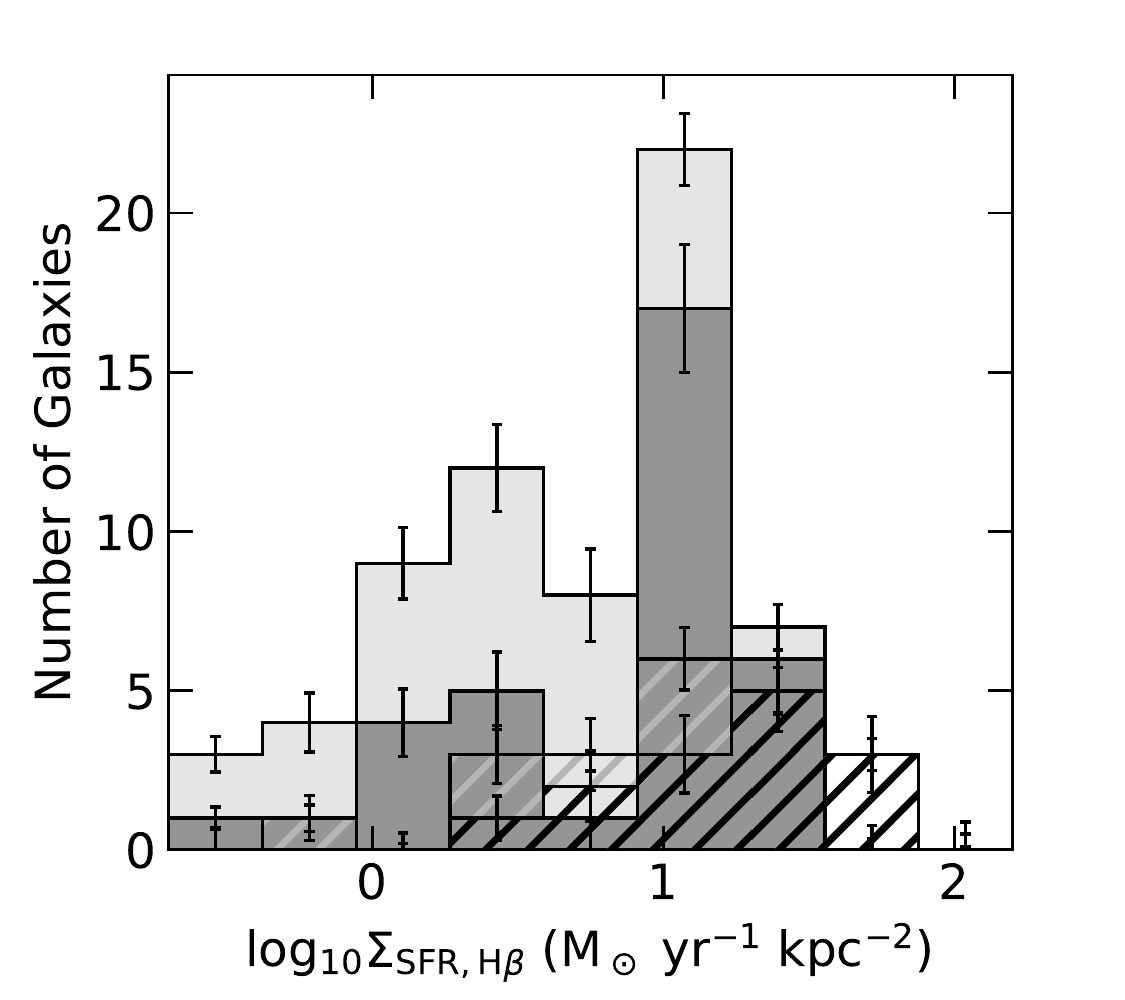}
    \caption{Same as Figure \ref{fig:ewhb_hist} but for \sigsfr\ derived from H$\beta$.\label{fig:sfr_hist}}% (top) and $F(1100)$ (bottom).}
\end{figure}

%We calculate the physical half-light radius $r_{50,NUV}$ and the extinction-corrected H$\beta$ and 1100 \AA\ luminosities, respectively. 
We convert the observed properties into a star formation rate surface density, \sigsfr, assuming the H$\beta$ and FUV star formation rate calibrations from \citet{2012ARA&A..50..531K} and dividing by $2\pi r_{50,NUV}^2$ following \citet{2020ApJ...892..109N}. { We note that these star formation rate indicators using H$\beta$ and UV luminosities are based on standard calibrations rather than tailored to the detailed properties (e.g., metallicity) of our sample. Thus, these star formation rates are more representative of the H$\beta$ and FUV luminosities than the true star formation rate. Figure \ref{fig:sfr_hist} shows that the LzLCS spans a much wider range in \sigsfr\ than previous studies. While many LzLCS LCEs exhibit \sigsfr$>10$ M$_\odot$ yr$^{-1}$ like their published LCE counterparts, many LzLCS LCEs have much lower \sigsfr. Since the LzLCS LCEs' half-light radii are similar to those of published LCEs, the difference in LCE \sigsfr\ is a distinction in SFR, suggesting concentration is more important than SFR for LyC escape.}

\begin{figure}
    \centering
    \includegraphics[width=\columnwidth]{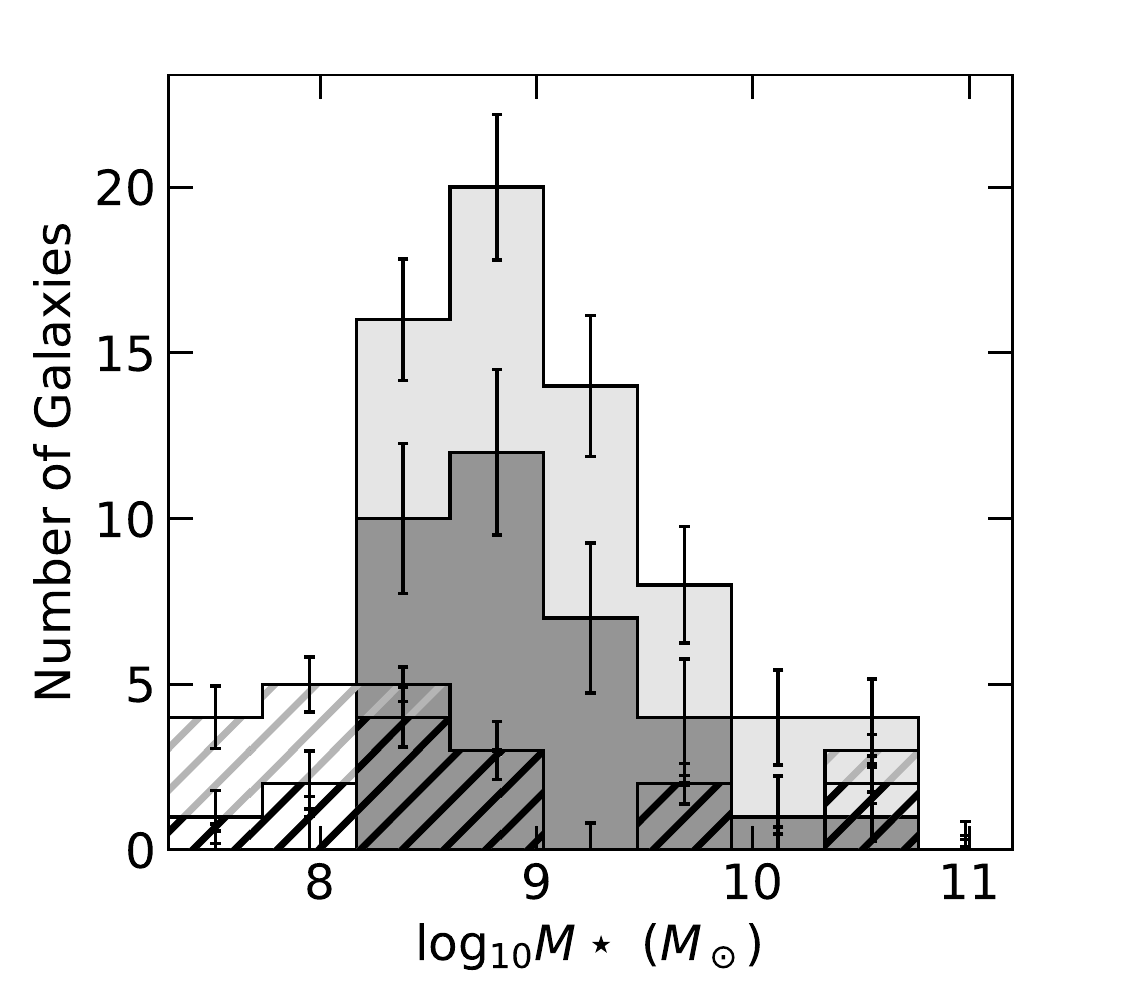}
    \caption{Same as Figure \ref{fig:ewhb_hist} but for $M_\star$.\label{fig:mstar_hist}}
\end{figure}

Stellar masses, $M_\star$, are determined by using the stellar population inference code {\sc Prospector} \citep{2017ApJ...837..170L,2019ascl.soft05025J} to fit aperture-matched photometry from SDSS and {\it GALEX} assuming a non-parametric star formation history, a \citet{2001MNRAS.322..231K} initial mass function, {\sc Cloudy} photoionization models, and a \citet{2001PASP..113.1449C} dust attenuation curve (Ji et al. in prep, see also Rutkowski et al. in prep), noting that the inferred stellar mass changes negligibly if adopting the \citet{2016ApJ...828..107R} extinction law instead. As shown in Figure \ref{fig:mstar_hist}, $M_\star$ ranges from $10^{8.25}$ to $10^{10.75}$ $M_\odot$ with half of the sample having $M_\star<10^9$ $M_\odot$. { The LzLCS LCEs are primarily dwarf galaxies but do persist up to much higher mass ($M_\star>10^{10}$ $M_\odot$) galaxies. While the distribution of LCEs suggests that dwarf galaxies dominate the LCE population, higher mass galaxies can still be LCEs.}

We use these stellar masses to compute the specific star formation rate, sSFR$=$SFR$_{{\rm H}\beta}/M_\star$.

\subsection{Ly$\alpha$ Escape Fraction}\label{sec:lyaesc}

\begin{figure}
    \centering
    \includegraphics[width=\columnwidth]{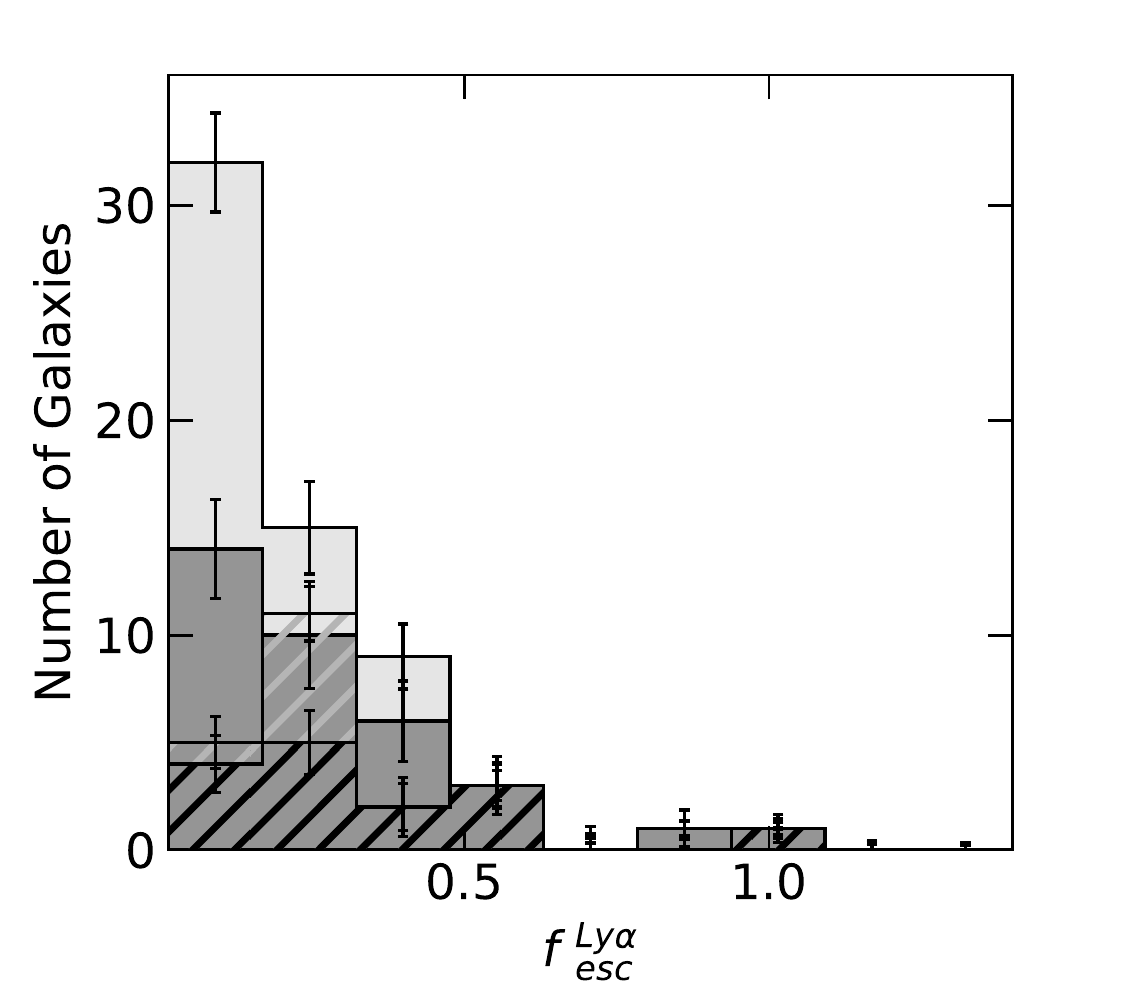}
    \caption{Same as Figure \ref{fig:ewhb_hist} but for \fesclya.\label{fig:lya_esc_hist}}
\end{figure}

Using the temperatures and densities derived from the optical emission lines, we compute the Case B emissivities for Ly$\alpha$ and H$\beta$ from the {\sc pyneb} grid of recombination coefficients from \citet{1995MNRAS.272...41S}. 
We correct H$\beta$ for both Galactic and internal extinction and use the ratio of Ly$\alpha$ to H$\beta$ emissivities to infer the intrinsic Ly$\alpha$ flux. We then use the observed Ly$\alpha$ flux, corrected for Galactic extinction, to calculate the fraction of Ly$\alpha$ photons, \fesclya, which escape the host galaxy. Uncertainties in the \fesclya\ are determined using Monte Carlo sampling of the grid of Case B emissivities from the uncertainties in temperature and density. We show the distribution of \fesclya\ values in Figure \ref{fig:lya_esc_hist}. While the LzLCS samples roughly the same range of \fesclya$\in[0,0.6]$ as previous studies, half of the sample exhibits low ($<16$\%) Ly$\alpha$ escape. This concentration at low \fesclya\ suggests a high \ion{H}{1} column along the line of sight in many of the LzLCS galaxies \citep[e.g.,][]{2015A&A...578A...7V}.

As demonstrated in Figure \ref{fig:lya_esc_hist}, two objects exhibit atypically high \fesclya: J081112+414146 and J164849+495751. These galaxies also have H$\alpha$/H$\beta$ and H$\gamma$/H$\beta$ decrements which are not permitted by traditional Case B limits of $\geq2.747$ and $\leq0.475$, respectively. { This inconsistency most likely indicates that the Balmer lines in SDSS spectra are problematic. Alternatively, such high \fesclya\ values may indicate collisional excitation of the \ion{H}{1} $n=2$ state in galactic winds or a hot diffuse halo \citep[e.g.,][]{2021ApJ...906..104C} or excess Ly$\alpha$ scattered into the line of sight \citep[e.g.,][]{1996ApJ...466..831G}. Excepting these two extreme cases, the LzLCS LCEs tend to have \fesclya$<0.5$ like their published counterparts. However, the LzLCS non-LCEs tend to have much lower \fesclya\ ($<0.15$). Such \fesclya\ values demonstrate that the LzLCS more robustly samples the LCE population at high \ion{H}{1} column densities than previous studies and even suggests LyC photons can escape even when a substantial amount of neutral gas is present.}

\subsection{Ly$\alpha$ Peak Velocity Separation}

\begin{figure}
    \centering
    \includegraphics[width=\columnwidth]{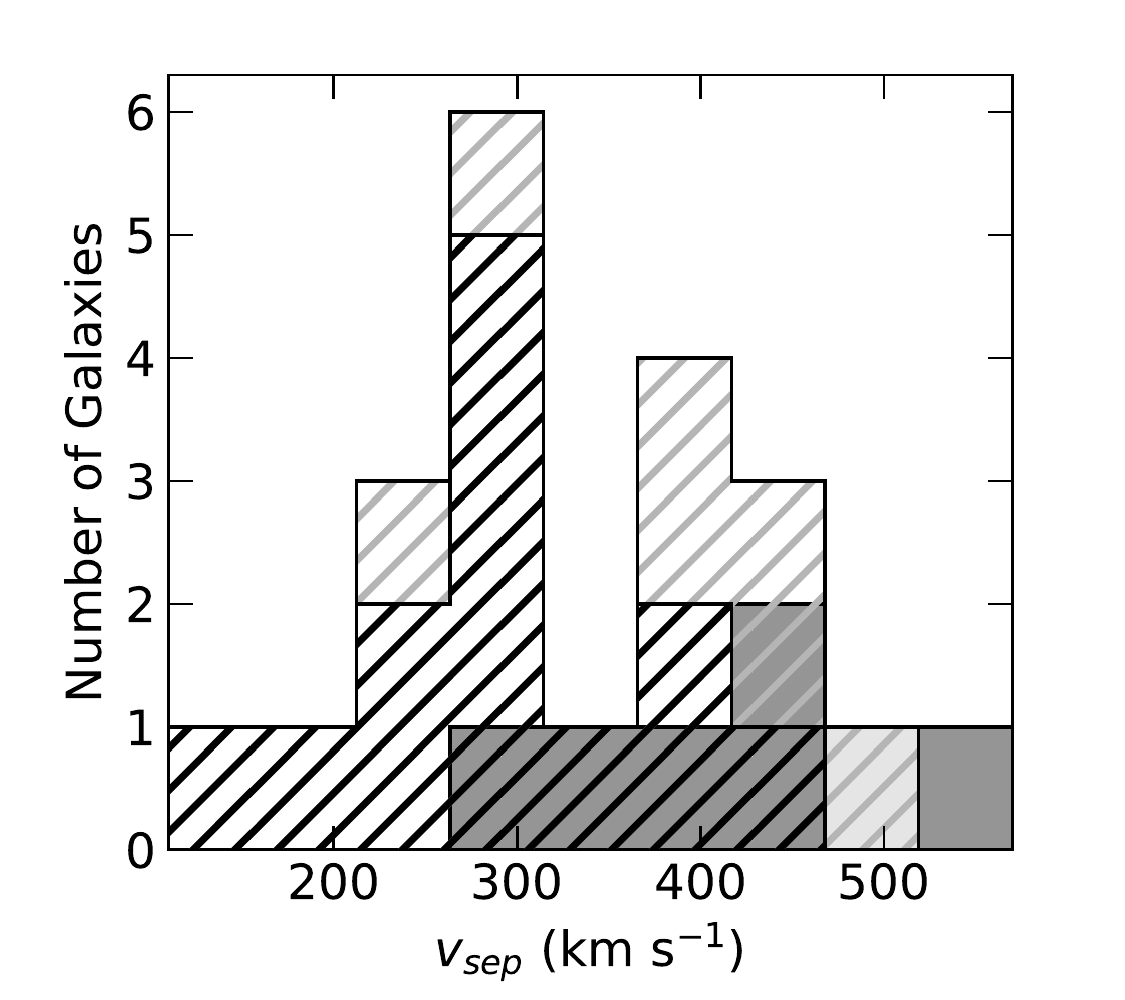}
    \caption{Same as Figure \ref{fig:ewhb_hist} but for the velocity separation of the blue and red peaks of the Ly$\alpha$ profile. No error bars are shown as insufficient uncertainties are reported in the literature. \label{fig:lya_peaksep}}
\end{figure}

The G140L resolution is insufficient to resolve the red and blue peaks of the Ly$\alpha$ profile. However, { seven targets} from the LzLCS have existing archival G160M COS spectra in which the two peaks are resolved. We obtain measurements of the velocity separation $v_{sep}$ of these peaks from \citet{2015ApJ...809...19H}, \citet{2017ApJ...844..171Y}, and \citet{2018A&A...616A..60O} and compare them to peak separations for published LCEs from \citet{2017A&A...597A..13V} and \citet{2018MNRAS.478.4851I} in Figure \ref{fig:lya_peaksep}. { While the number of LzLCS galaxies with measured $v_{sep}$ is small, the LCEs in our sample have Ly$\alpha$ peak separations larger than the characteristic value for published LCEs. This difference indicates a larger \ion{H}{1} column density in the LzLCS LCEs than in published LCEs, which may suggest LyC escape can occur in a variety of ISM geometries. However, the lack of $v_{sep}$ measurements prevents further insight.}

\subsection{{\it UV} Magnitudes}

%We correct the {\it GALEX} magnitudes for Galactic and internal reddening. To determine the extinction coefficients for the {\it GALEX} passbands, we integrate the extinction coefficients for an assumed reddening law weighted by the {\it GALEX} transmission functions for the FUV and NUV filters. For Galactic extinction, we use the same \citet{2018MNRAS.478..651G} reddening values and \citet{1999PASP..111...63F} law as described above. For internal extinction, we compute extinction coefficients by converting the {\it GALEX} transmission function to the rest frame. We then correct for internal extinction using $E(B-V)$ values derived from SED fitting of the UV continuum as described above. Using the {\it GALEX} transmission function, we subtract the contribution of Ly$\alpha$ flux to the FUV magnitude. Finally, we convert from apparent to absolute magnitudes using the distances obtained from the spectroscopic redshift and {\sc astropy}.

We compute the absolute UV magnitude at 1500 \AA\ (M$_{1500}$) from the {\sc Starburst99} templates best-fit to the COS spectra after correcting for Galactic (but \emph{not} internal) extinction. These magnitudes are computed by summing the template flux density over a 20 \AA\ boxcar window and converting to absolute AB magnitude using the luminosity distance derived from the spectroscopic redshift.

\begin{figure}
    \centering
    \includegraphics[width=\columnwidth]{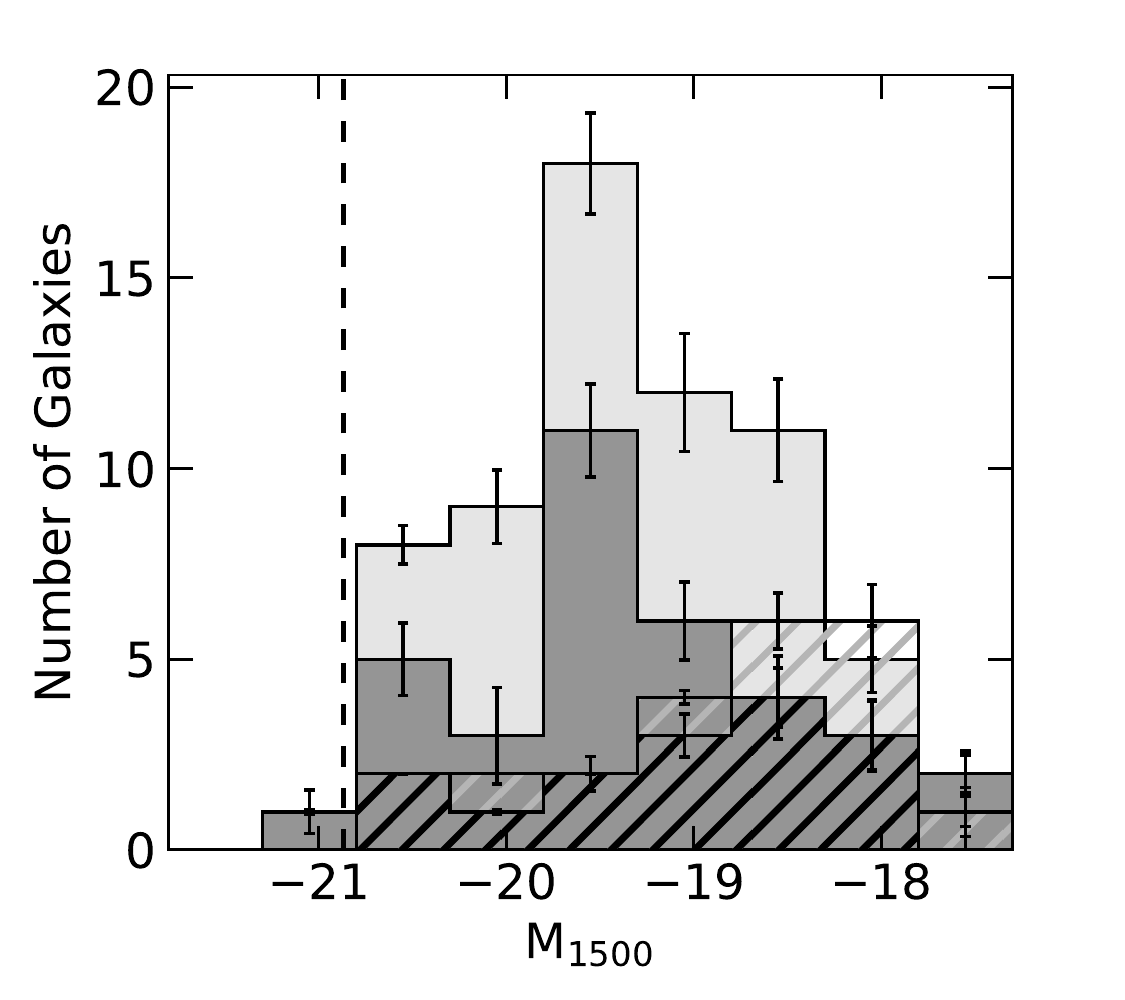}
    \caption{Same as Figure \ref{fig:ewhb_hist} but for M$_{1500}$. Dashed line indicates the characteristic $M_{1600}^*=-20.87$ at $z=7$ \citep{2015ApJ...803...34B}.\label{fig:muv_hist}}
\end{figure}

We show the COS UV magnitudes in Figure \ref{fig:muv_hist}, finding M$_{1500}\in[-22,-17]$. Figure \ref{fig:muv_hist} indicates that the younger stellar populations in many LzLCS galaxies are, like their published counter parts, { fainter than the characteristic $M^\star$. However, the LCEs in LzLCS span a wider range in $M_{1500}$ than the published LCEs, indicating that more luminous galaxies can also be LCEs.}%, with only a few galaxies populating the high luminosity tail of the magnitude distribution.
%We show the {\it GALEX} magnitudes in Figure \ref{fig:muv_hist}, finding M$_{FUV}\in[-26,-18.6]$ and M$_{NUV}\in[-24.8,-19.2]$. While both magnitude ranges represent more than two decades in UV luminosity, Figure \ref{fig:muv_hist} indicates that the majority of LzLCS galaxies are, like their published LCE counterparts, fainter galaxies, with only a few galaxies populating the high luminosity tail of the magnitude distributions. 

\section{LyC Escape Fraction\label{sec:fesc}}

We use three estimates of the LyC escape fraction, \fesclyc: the $F_{\lambda \rm LyC}/F_{\lambda 1100}$ flux ratio, \fesclyc\ derived from H$\beta$, and \fesclyc\ determined from fits to the UV continuum. { Each independent metric allows us to assess possible systematics in \fesclyc, providing an additional constraint on how much the escape fraction depends on our assumptions about, e.g., star formation history, dust extinction, etc.} The $F_{\lambda \rm LyC}/F_{\lambda 1100}$ flux ratio is an empirical proxy for \fesclyc\ \citep[cf.][]{2019ApJ...885...57W}. While less direct than \fesclyc, $F_{\lambda \rm LyC}/F_{\lambda 1100}$ is free of any assumptions about stellar populations or dust. { However, this flux ratio depends implicitly on extinction, burst age, and metallicity, making a direct interpretation less meaningful.} Values for this flux ratio span 0.0 to 0.328 with a median of 0.023.

\begin{figure*}
    \centering
    \includegraphics[width=\linewidth]{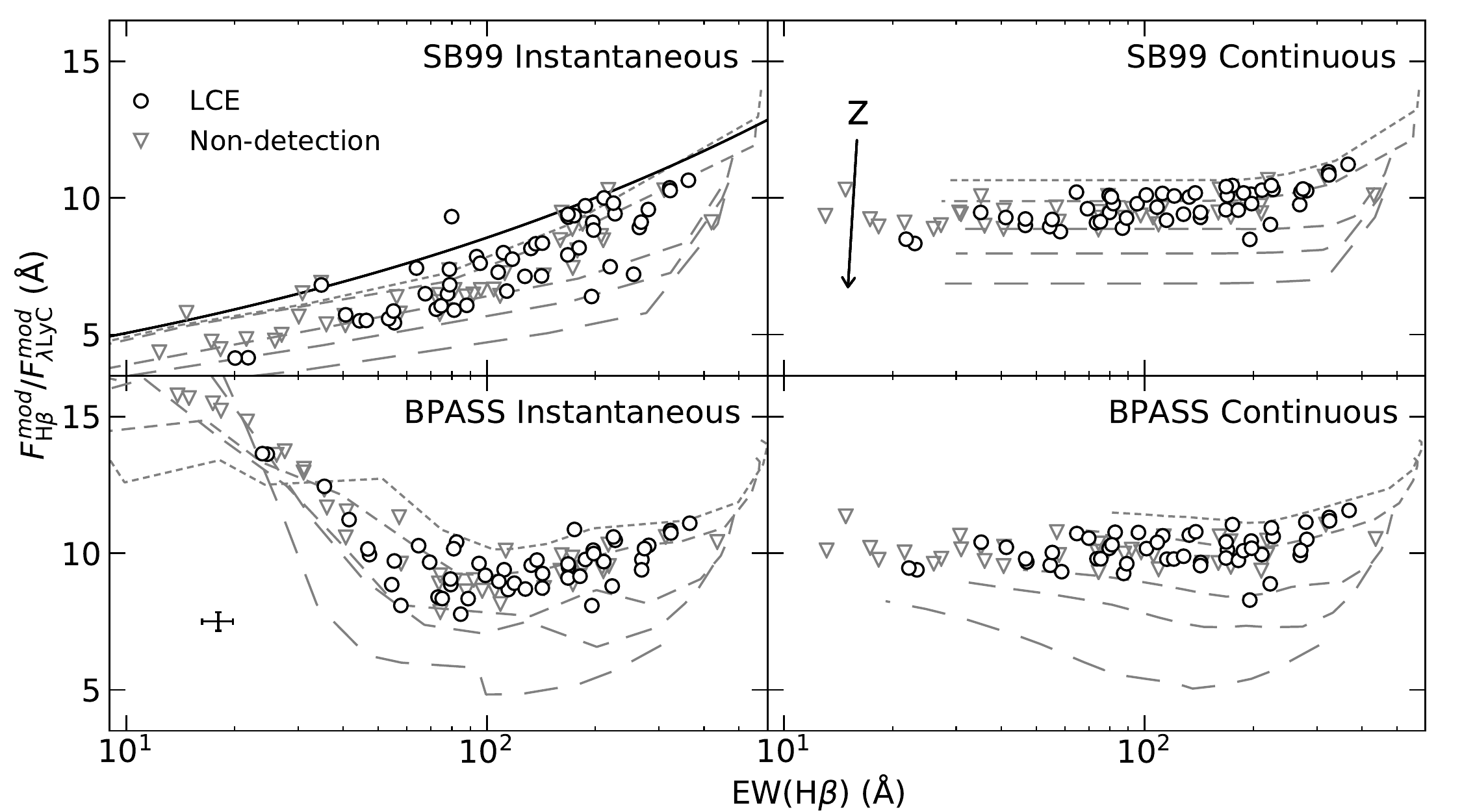}
    \caption{ $F_{\rm{H}\beta}^{mod}/F_{\lambda900}^{mod}$ ratio predicted by  instantaneous ({\it left}) and continuous ({\it right}) star formation histories from {\sc Starburst99} ({\it top}) and {\sc bpass} ({\it bottom}) models for \fesclyc$=0$ as a function of H$\beta$ EW for a range of metallicities. Metallicity increases with increasing dash length over the interval $Z=[0.001,0.04]$, as shown by the arrow in the top right panel. { The change in $F_{{\rm H}\beta}/F_{\lambda\rm LyC}$ with metallicity is due to the softening of stellar SEDs with increasing metallicity.} Solid black line ({\it top left}) indicates the \citet{2016MNRAS.461.3683I} relation. Symbols represent the inferred $F_{\rm{H}\beta}/F_{\lambda\rm LyC}$ for \fesclyc$=0$ for LCEs (circles) and non-detections (open triangles) implied by the measured $F_{\lambda\rm LyC}$, H$\beta$, $12+\log_{10}\left(\frac{\rm O}{\rm H}\right)$, and \fesclyc\ for each set of models. Median uncertainties are shown in the lower left panel, increased by a factor of three for visualization.\label{fig:synthpop_models}}
\end{figure*}

\begin{figure}
    \centering
    \includegraphics[width=\columnwidth]{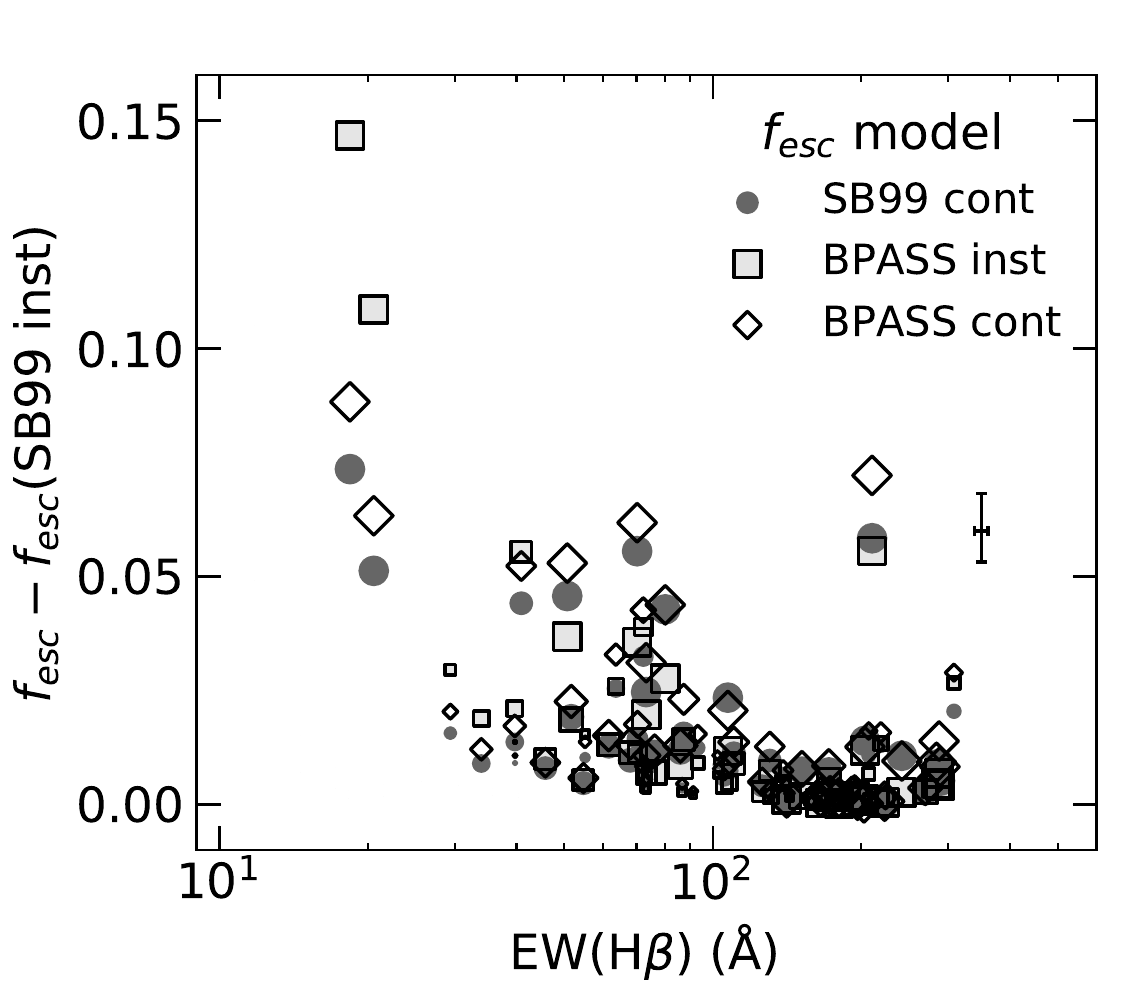}
    \caption{ Difference between \fesclyc\ calculated from {\sc bpass} instantaneous (open diamonds) and continuous (light grey squares) star formation or from {\sc Starburst99} continuous star formation (dark grey circles) models and \fesclyc\ calculated using {\sc Starburst99} instantaneous star formation models as compared to H$\beta$ EW for \fesclyc\ derived from H$\beta$ for the combined LzLCS and published samples. Symbol size corresponds to the LyC detection significance. Characteristic uncertainties are shown on the right. Increasing scatter with decreasing H$\beta$ EW indicates the increasing effect of assumptions about star formation history on the inferred \fesclyc. \label{fig:fesc_age}}
\end{figure}

\subsection{H$\beta$ \fesclyc}

To estimate the absolute \fesclyc, we use the extinction-corrected flux and rest-frame EW of H$\beta$ to infer the LyC absorbed by the ISM in the galaxy as described by \citet{2016MNRAS.461.3683I,2018MNRAS.474.4514I}. { This approach assumes that the extinction-corrected H$\beta$ flux is a proxy for the total number of ionizing photons absorbed by the nebula \citep[e.g.,][]{source:osterbrock2006}. The conversion between the total ionizing photon flux and the LyC flux at a particular wavelength depends on the stellar population age. The H$\beta$ EW yields the burst age for an assumed star formation history. Use of the H$\beta$ EW is necessary to select the appropriate model $F_{\rm{H}\beta}^{mod}/F_{\lambda \rm LyC}^{mod}$ independent of the UV SED fits. Here, the $F_{\rm{H}\beta}^{mod}/F_{\lambda \rm LyC}^{mod}$ ratio accounts for the shape of the LyC by tracing the amount of ionizing flux that falls within the 20 \AA\ bin over which the LyC flux is measured.} Because the ionization cross-section of \ion{H}{1} is proportional to $\lambda^3$, the emergent flux increases with decreasing LyC wavelength for a fixed escape fraction. We calculate the synthetic LyC flux in the 20\ \AA\ bins used to measure the LyC in the COS spectra, matching the model and observed wavelength bins.

{ To obtain model LyC and H$\beta$ values for deriving \fesclyc, we consider two sets of population models, each with two different star formation histories. To predict LyC fluxes and H$\beta$ properties, we use {\sc Starburst99} models \citep{1999ApJS..123....3L,2014ApJS..212...14L} and {\sc bpass} models \citep{2018MNRAS.479...75S} assuming either an instantaneous burst or continuous star formation.} Model ages span from 0.1 to 500 Myr while metallicities range from $Z=0.001$ to $0.04$. For the {\sc Starburst99} and {\sc bpass} models, we assume a \citet[]{2001MNRAS.322..231K} and \citet[]{2003PASP..115..763C} initial mass functions, respectively, with mass ranges from 0.1 to 100 M$_\odot$ for both sets of models. In all cases, the shape of the relation changes as a function of stellar metallicity (see Figure \ref{fig:synthpop_models}), meaning there is no single function from which to infer the absorbed LyC flux. To select the appropriate sequence, we take advantage of our $12+\log_{10}({\rm O/H })$ abundance estimate derived in \S \ref{sec:nebabn}, assuming that the gas and stellar metallicities are comparable since the stars have recently formed and have not yet further enriched the ISM. { While at high redshift, $\alpha$ enhancement may affect the scaling of oxygen to total metalicity; however, \citet{2011ApJ...728..161I} find that GPs have Fe/O ratios comparable to those of other low-redshift dwarf galaxies and that dust depletion sufficiently explains any apparent gas-phase $\alpha$ excess relative to iron. Thus, scaling the total metallicity by the relative oxygen abundance is appropriate for the combined sample.}

{ We then interpolate over the grid of $F_{{\rm H}\beta}/F_{\lambda \rm LyC}$ from the stellar population models to obtain the predicted intrinsic flux ratio vs H$\beta$ EW for the given galaxy's metallicity.} As in \citet{2016MNRAS.461.3683I}, we take the LyC flux $F_{\lambda \rm LyC}^{abs}$ implied by the H$\beta$ flux to be the absorbed LyC flux such that the \fesclyc\ is
\begin{equation}
    f_{esc}^{LyC}({\rm H}\beta) = \frac{F_{\lambda \rm LyC}^{obs}}{F_{\lambda \rm LyC}^{obs}+F_{\lambda \rm LyC}^{abs}} = \frac{F_{\lambda \rm LyC}^{obs}}{F_{\lambda\rm LyC}^{mod}}.
\end{equation}
{ This relation is only an initial estimate as the H$\beta$ EW, the burst age indicator used to infer $F_{\rm{H}\beta}^{mod}/F_{\lambda \rm LyC}^{mod}$,} is in fact affected by LyC escape, meaning that H$\beta$ EW must be corrected for $f_{esc}$ in order to yield the appropriate flux ratio \citep[e.g.,][]{2018MNRAS.478.4851I}. Using the \fesclyc\ calculated from the uncorrected value as the initial condition, we iteratively correct H$\beta$ EW and recompute $f_{esc}$ until converging on a value of \fesclyc, typically within 10 or fewer iterations. Uncertainties in \fesclyc\ are estimated by Monte Carlo simulation, sampling the uncertainties in H$\beta$ flux and EW, LyC flux, and $12+\log_{10}({\rm O/H })$ and recalculating \fesclyc\ $10^4$ times. 

{As evident in Figure \ref{fig:synthpop_models}, the {\sc Starburst99} models reproduce the \citet{2016MNRAS.461.3683I} relation between H$\beta$ EW and the $F_{\rm{H}\beta}/F_{\lambda \rm LyC}$ flux ratio for an instantaneous starburst of 10\% solar metallicity (Figure \ref{fig:synthpop_models}, upper left). However, our results indicate that using the \citet{2016MNRAS.461.3683I} prescription will consistently yield higher $F_{\rm{H}\beta}/F_{\lambda \rm LyC}$ and thus over-estimate \fesclyc\ for an instantaneous burst.} { We also take into account the effects of binary star evolution by considering the {\sc bpass} model \citep{2018MNRAS.479...75S}. This model increases the ionizing photon budget by an amount comparable to that of the continuous starburst (Figure \ref{fig:synthpop_models}, lower left).  Continuous {\sc Starburst99} and {\sc BPASS} models predict $F_{\rm{H}\beta}/F_{\lambda \rm LyC}$ values higher than those of the {\sc Starburst99} instantaneous burst models (Figure \ref{fig:synthpop_models}, upper and lower right), with the difference increasing as H$\beta$ EW decreases due to subsequent generations of young stars.}

{ Using the H$\beta$ line with the {\sc Starburst99} and {\sc BPASS} continuous star formation models yields \fesclyc\ values ranging from 0 to 20\% for the LzLCS sample and from 0 to 45\% for the published LCEs. 15 of the LzLCS LCEs and 9 of the published LCEs have cosmologically relevant values of \fesclyc$>0.05$. The {\sc BPASS} instantaneous burst models yield similar results, with a median difference in \fesclyc\ of $\approx0.001$\%. The \fesclyc\ values derived from {\sc Starburst99} instantaneous star formation models diverge from the {\sc bpass} and {\sc Starburst99} \fesclyc\ with increasing burst age.} We illustrate this effect in Figure \ref{fig:fesc_age} by comparing the difference in \fesclyc\ to the H$\beta$ EW. For the youngest bursts, the difference is negligible because the early O stars dominate the LyC and optical continuum in every scenario; however, the effects of continuously forming new O and B stars or accretion onto stripped stars in binary star systems can amplify \fesclyc\ by as much as a factor of two or three, respectively, at later burst ages.

\begin{figure*}
    \centering
    \includegraphics[width=0.4965\linewidth]{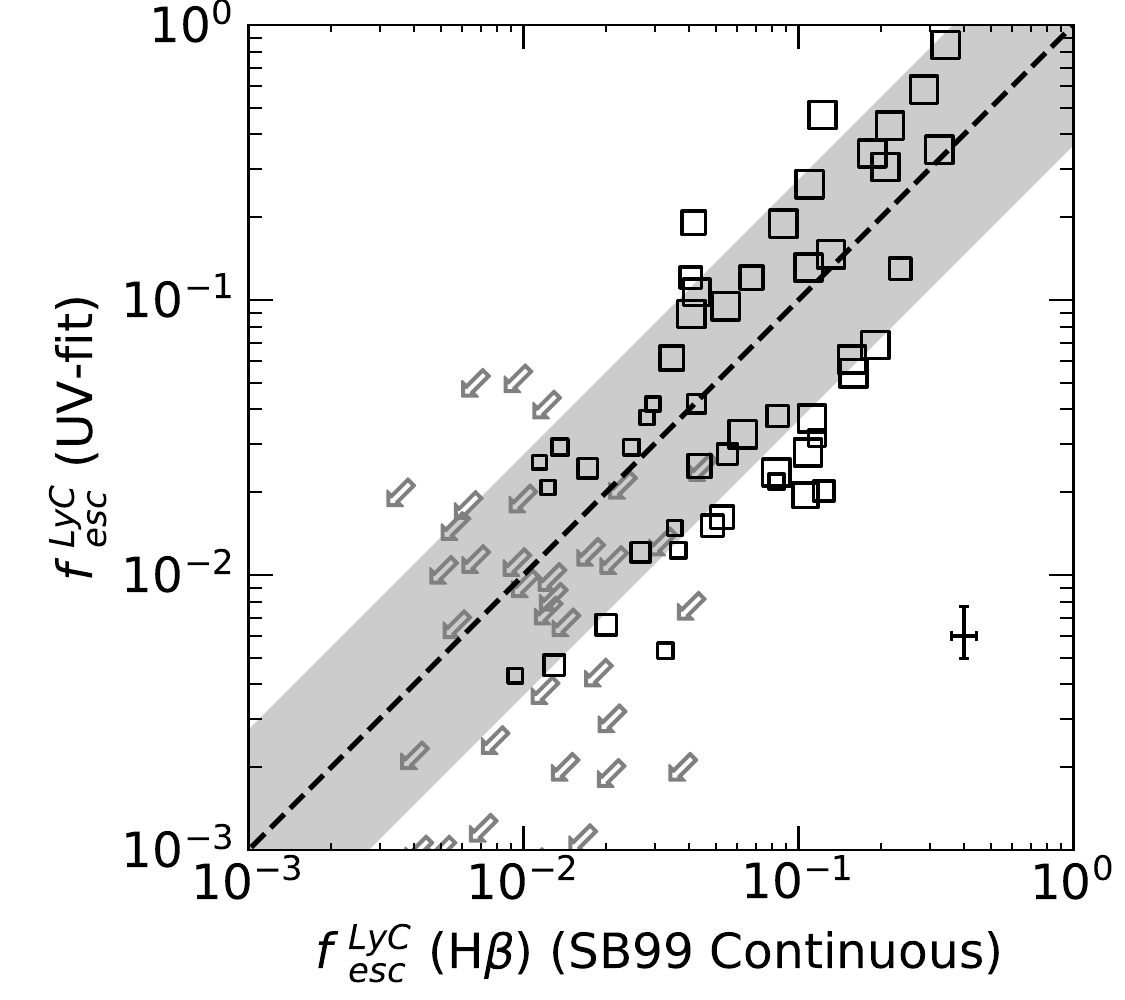}
    \includegraphics[width=0.4965\linewidth]{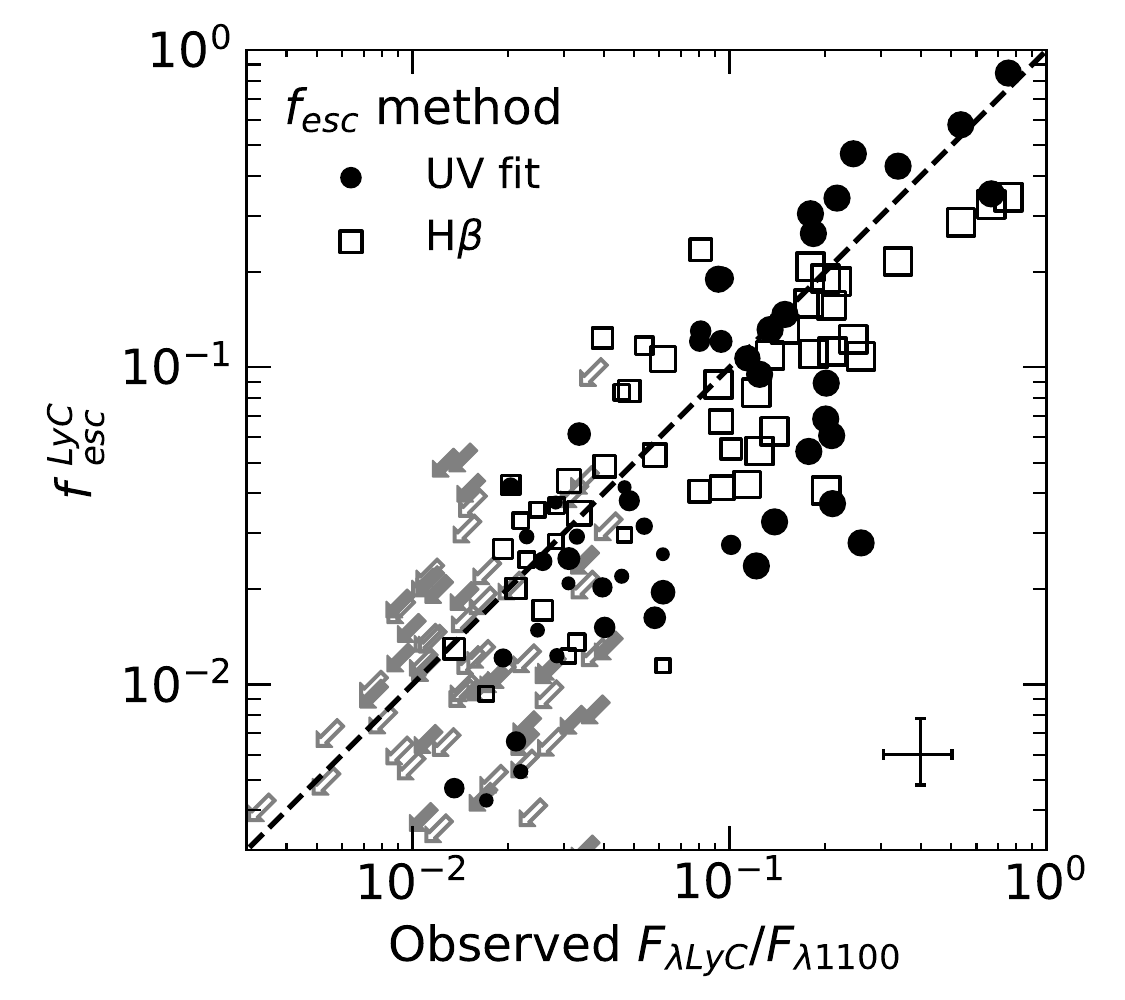}
    \caption{{\it Top}: \fesclyc\ derived from fitting the COS UV spectrum compared to \fesclyc\ derived from H$\beta$ using the {\sc Starburst99} continuous star formation models. Shaded region indicates the rms log difference of the two \fesclyc\ values. {\it Bottom}: Comparison of \fesclyc\ with $F_{\lambda \rm LyC}/F_{\lambda 1100}$ for the combined sample of LzLCS and published LCEs. Open symbols represent \fesclyc\ derived from H$\beta$ using the {\sc Starburst99} continuous star formation models while filled symbols represent \fesclyc\ derived from fits to the UV continuum. In both figures, symbol size corresponds to the LyC detection significance, and grey arrows indicate upper limits on \fesclyc. Error bars in the lower right indicate the median uncertainty for objects with detected LyC ($P(>N|B)<0.02275$). Dashed line is 1:1 agreement.\label{fig:fesc_flycf1100}}
\end{figure*}

\subsection{UV SED \fesclyc}

To determine \fesclyc\ from the UV continuum fits, we compute the ratio of the measured LyC flux to the intrinsic (unreddened) LyC flux implied by the best-fit {\sc Starburst99} templates. We obtain the intrinsic LyC flux by using the low-resolution {\sc Starburst99} bases, summing the model flux in the same spectral window used to measure the LyC in the COS spectrum. The UV escape fraction is then obtained by
\begin{equation}
f_{esc}^{LyC}({\rm UV}) = \frac{F_{\lambda \rm LyC}^{obs}}{F_{\lambda \rm LyC}^{fit}}.
\end{equation}
This yields a range of \fesclyc\ from 0 to 50\% with 9 of the LzLCS LCEs having \fesclyc$>0.05$. While the uncertainties for the UV \fesclyc\ are higher due to the SNR of the COS spectrum, this approach is less sensitive to the presence of older stellar populations than H$\beta$ because the FUV continuum is only sensitive to the youngest stars.

\subsection{Comparison of \fesclyc\ Results}

{ In Figure \ref{fig:fesc_flycf1100}, we show that the UV and H$\beta$ methods typically agree to within $\sim$0.5 dex, which we confirm by calculating the rms ratio of the two \fesclyc\ values. Even the strongest LCEs exhibit this scatter, indicating the persistence of systematic uncertainty across a dynamic range of escape fractions.} The UV \fesclyc\ values are consistently higher at high values of $F_{\lambda \rm LyC}/F_{\lambda 1100}$. While both approaches to deriving \fesclyc\ depend on assumptions, the consistency between all three measures of LyC escape gives us confidence in assessing the relevance of different galaxies to reionization. {Despite the effects of extinction and stellar populations implicit to the measured $F_{\lambda \rm LyC}/F_{\lambda 1100}$, we show in Figure \ref{fig:fesc_flycf1100} that \fesclyc\ correlates well with it, indicating this flux ratio is a rough proxy for \fesclyc. Of all the H$\beta$ \fesclyc\ estimates, the {\sc Starburst99} continuous star formation models yield \fesclyc\ values most consistent with the UV \fesclyc. { This agreement suggests the the UV and continuous H$\beta$ EW star formation histories are most comparable, although discrepancies still persist between the two.} We find that the observed $F_{\lambda \rm LyC}/F_{\lambda 1100}$ are best predicted by the continuous star formation models. Thus, we proceed with \fesclyc\ derived using the continuous star formation models in subsequent analysis.}

{While the UV-fit \fesclyc\ tends to be higher at high $F_{\lambda \rm LyC}/F_{\lambda 1100}$, only 9 LzLCS targets have significantly high \fesclyc\ ($>0.05$). Despite lower maximum \fesclyc\ values, H$\beta$ yields 15 objects with high \fesclyc. In our re-measurement of \fesclyc\ for published {\it HST}/COS LyC observations, we find 13 have high \fesclyc\ from fits to the UV continuum and 14 have high \fesclyc\ from H$\beta$.} {Thus, through the LzLCS, we have roughly doubled the number of known LCEs with cosmologically significant \fesclyc\ ($\gtrsim5$ \%), demonstrating the immense scientific value of the LzLCS program.}

\section{Conclusion}

We present the Low-redshift Lyman Continuum Survey, the largest search for LCEs in the low-redshift ($z\sim0.3$) universe. With careful processing of {\it HST}/COS spectra, we measure the LyC in 66 candidate LCEs, detecting flux with $>$97.725\% significance from 35 galaxies in the sample. The LzLCS nearly triples the number of known local LCEs.

From UV and optical spectra and UV photometry, we characterize the global properties of the LzLCS galaxies. The sample contains low metallicity galaxies with direct-method oxygen abundances ranging from $12+\log_{10}({\rm O/H })$=7.5 to 8.5, a much broader range than previously published LCEs. Stellar masses span $M_\star=10^{8}$ to $10^{10}$ $M_\odot$ with the UV halflight radii ranging from 0.3 to 2.25 kpc. %The UV $\beta$ slope, \fesclya, UV ISM absorption lines, Balmer decrements indicate a large range of extinction and covering fractions. 
The halflight radii, \sigsfr, and $sSFR$\ imply highly concentrated star formation. H$\beta$ EWs and \orat\ imply a range of starburst ages, ionization parameter, and/or optical depth effects. { {The LzLCS covers a wider range of properties than previously published low-redshift LCEs, demonstrating the ability of our survey to explore and test the heterogeneity of LCEs and \fesclyc.}}

From empirical methods and synthetic stellar population models, we derive escape fractions ranging from 0 to 50\%. Although previous studies suggest \fesclyc\ derived by different methods agree \citep[e.g.,][]{2016MNRAS.461.3683I,2018MNRAS.478.4851I}, the broader scope of the LzLCS demonstrates that \fesclyc\ estimates can be sensitive to assumptions about stellar populations and star formation history as well as the data used. Based on our assessment of different methods and related systematic uncertainties, the \fesclyc\ based on the UV starlight continuum is the most reliable because it is less sensitive to assumptions than the \fesclyc\ based on H$\beta$.

With the LzLCS, we have roughly doubled the number of local LCEs with cosmologically relevant LyC escape \citep[\fesclyc$>0.05$, e.g.,][]{2015ApJ...802L..19R} from 13 to 22 (or 14 to 29) using the UV (or H$\beta$) method. The LzLCS thus offers an unprecedented opportunity to investigate the conditions related to LyC escape in galaxies, the results of which may be extended to the epoch of reionization.

We evaluate LCE and \fesclyc\ diagnostics in companion papers \citep[Flury et al. submitted,][]{2021arXiv210403432W}. We also analyze the UV absorption lines (Salda\~na-Lopez et al. submitted) and SED parameters (Ji et al. in prep) of LCEs and non-emitters. Additional planned work includes investigation of the shape of the LyC and the Lyman break, feedback and gas dynamics, neutral and low-ionization gas covering fractions, and photoionization modeling to further understand the ISM conditions and physical mechanisms of escaping LyC.

\begin{acknowledgments}

We thank the anonymous referee for feedback which improved the clarity of this paper.

Support for this work was provided by NASA through grant number \emph{HST}-GO-15626 from the Space Telescope Science Institute. Additional work was based on observations made with the NASA/ESA Hubble Space Telescope, obtained from the data archive at the Space Telescope Science Institute from \emph{HST} proposals 13744, 14635, 15341, and 15639. STScI is operated by the Association of Universities for Research in Astronomy, Inc. under NASA contract NAS 5-26555.

Funding for the Sloan Digital Sky 
Survey IV has been provided by the 
Alfred P. Sloan Foundation, the U.S. 
Department of Energy Office of 
Science, and the Participating 
Institutions. SDSS-IV acknowledges support and 
resources from the Center for High 
Performance Computing  at the 
University of Utah. The SDSS 
website is \url{www.sdss.org}.
SDSS-IV is managed by the 
Astrophysical Research Consortium 
for the Participating Institutions 
of the SDSS Collaboration.

RA acknowledges support from ANID Fondecyt Regular 1202007.

\end{acknowledgments}

\software{{\sc\ astropy\ } \citep{astropy:2013,astropy:2018}, {\sc\ bpass\ } \citep{2018MNRAS.479...75S}, {\sc\ calcos}, {\sc\ Cloudy\ } \citep{2013RMxAA..49..137F}, {\sc\ emcee\ } \citep{2013PASP..125..306F}, {\sc\ FaintCOS\ } \citep{2016ApJ...825..144W,2020arXiv201207876M}, {\sc\ matplotlib\ } \citep{matplotlib}, {\sc\ numpy\ } \citep{numpy}, {\sc\ Prospector\ } \citep{2017ApJ...837..170L,2019ascl.soft05025J}, {\sc\ pyneb\ } \citep{2015A&A...573A..42L}, {\sc\ scipy\ } \citep{scipy}, {\sc Starburst99\ } \citep{1999ApJS..123....3L,2010ApJS..189..309L,2014ApJS..212...14L}}

\newpage
\bibliographystyle{aasjournal}
\bibliography{biblio}

\startlongtable
\begin{deluxetable*}{lccccc}
\tablecaption{Flux ratios measured from the SDSS optical spectra for the combined LzLCS and \pubsamp\ samples. A full version of this table is available online. \label{tab:optical_frat}}
\tablewidth{\textwidth}
\tablehead{
\colhead{Object} & \colhead{EW(H$\beta$)} & \colhead{$\log_{10}R_{23}$} & \colhead{$\log_{10}O_{32}$} & \colhead{$\log_{10}O_{31}$} & \colhead{$\log_{10}$[O I]/H$\beta$} \\
\colhead{} & \colhead{\AA}
}
\startdata
J003601+003307 &  160.469$\pm$~6.509 &  ~0.938$\pm$~0.024 &  ~1.113$\pm$~0.039 &          $>$~2.621 &          $<$-1.836 \\
J004743+015440 &   61.511$\pm$~1.137 &  ~0.956$\pm$~0.024 &  ~0.655$\pm$~0.026 &          $>$~2.870 &          $<$-2.124 \\
J011309+000223 &   40.896$\pm$~1.775 &  ~0.964$\pm$~0.076 &  ~0.357$\pm$~0.086 &          $>$~2.349 &          $<$-1.668 \\
J012217+052044 &   87.255$\pm$~3.110 &  ~0.942$\pm$~0.041 &  ~0.881$\pm$~0.047 &          $>$~2.553 &          $<$-1.790 \\
J012910+145935 &   73.370$\pm$~1.180 &  ~0.855$\pm$~0.026 &  ~0.340$\pm$~0.031 &  ~1.717$\pm$~0.088 &  -1.150$\pm$~0.088
\enddata
\end{deluxetable*}

\startlongtable
\begin{deluxetable*}{lcccc}
\tablecaption{Properties derived from the SDSS optical spectra for the combined LzLCS and \pubsamp\ samples. A full version of this table is available online.\label{tab:optical_props}}
\tablewidth{\textwidth}
\tablehead{
\colhead{Object} & \colhead{$\log_{10}$SFR$_{\mathrm{H}\beta}$} & \colhead{$n_e$ (cm$^{-3}$)} &\colhead{$T_e$ (K)} & \colhead{$12+\log_{10}\left(\frac{\rm O}{\rm H}\right)$}
}
\startdata
J003601+003307 & ~1.184$\pm$~0.024 &          100 &       15770$\pm$970 &      ~7.781$\pm$~0.037 \\
J004743+015440 & ~1.336$\pm$~0.024 & 1610$\pm$150 &       13820$\pm$850 &      ~8.290$\pm$~0.037 \\
J011309+000223 & ~0.699$\pm$~0.076 & 6590$\pm$180 &      11640$\pm$1060 &      ~8.329$\pm$~0.115 \\
J012217+052044 & ~0.971$\pm$~0.041 &          100 &      16220$\pm$1760 &      ~7.799$\pm$~0.064 \\
J012910+145935 & ~1.125$\pm$~0.026 & 2600$\pm$120 &       10000$\pm$250 &      ~8.411$\pm$~0.044
\enddata
\end{deluxetable*}

\startlongtable
\begin{deluxetable*}{lccccccc}
\tablecaption{Properties derived from the {\it HST}/COS G140L spectra for the combined LzLCS and\\ \pubsamp\ samples. A full version of this table is available online.\label{tab:uv_props}}
\tablewidth{\textwidth}
\tablehead{
\colhead{Object} & \colhead{$M_{1500}$} & \colhead{$r_{50}$} & \colhead{$\beta_{1200}$} & \colhead{$f_{1100}\times10^{17}$} & \colhead{EW(Ly$\alpha$)} &  \colhead{$\log_{10}$SFR$_{f_{1100}}$} \\
\colhead{} & \colhead{} & \colhead{kpc} & \colhead{} & \colhead{erg s$^{-1}$ cm$^{-2}$ \AA$^{-1}$} & \colhead{\AA} & \colhead{$M_\odot$ yr$^{-1}$}
}
\startdata
  J003601+003307 &   -17.905$\pm$~0.100 &  ~0.445$\pm$~0.148 & -2.900$\pm$~0.328 &  ~6.445$_{-0.381}^{+0.418}$ &        93.900$\pm$~9.330 &                    ~0.345$\pm$~0.028 \\
  J004743+015440 &   -20.300$\pm$~0.094 &  ~0.618$\pm$~0.145 & -2.380$\pm$~0.298 &  38.611$_{-2.507}^{+2.787}$ &        41.526$\pm$~4.427 &                    ~1.325$\pm$~0.031 \\
  J011309+000223 &   -19.657$\pm$~0.118 &  ~0.627$\pm$~0.133 & -1.990$\pm$~0.253 &  36.421$_{-3.649}^{+2.623}$ &        31.291$\pm$~3.560 &                    ~1.252$\pm$~0.031 \\
  J012217+052044 &   -19.410$\pm$~0.098 &  ~0.713$\pm$~0.151 & -1.609$\pm$~0.274 &  23.193$_{-1.595}^{+1.427}$ &        70.616$\pm$~6.793 &                    ~0.922$\pm$~0.027 \\
  J012910+145935 &   -19.862$\pm$~0.058 &  ~0.636$\pm$~0.127 & -1.672$\pm$~0.205 &  50.705$_{-3.600}^{+2.979}$ &        39.593$\pm$~4.840 &                    ~1.370$\pm$~0.026
\enddata
\end{deluxetable*}

\startlongtable
\begin{deluxetable*}{lcccc}
\tablecaption{Properties derived jointly from {\it HST}/COS, and SDSS data for the combined LzLCS and \pubsamp\ samples. A full version of this table is available online.\label{tab:joint_props}}
\tablewidth{\textwidth}
\tablehead{
\colhead{Object} & \colhead{$f_{esc}^{\mathrm{Ly}\alpha}$} & \colhead{$\log_{10}\Sigma_{SFR,\mathrm{H}\beta}$} & \colhead{$\log_{10}\Sigma_{SFR,f_{1100}}$} & \colhead{$\log_{10}M_\star$}
}
\startdata
  J003601+003307 &                       ~0.116$\pm$~0.011 &                                 ~1.880$\pm$~0.146 &                          ~0.249$\pm$~0.147 &   ~8.754$_{-0.425}^{+0.444}$ \\
  J004743+015440 &                       ~0.194$\pm$~0.019 &                                 ~0.956$\pm$~0.105 &                          ~0.945$\pm$~0.107 &   ~9.203$_{-0.430}^{+0.439}$ \\
  J011309+000223 &                       ~0.398$\pm$~0.075 &                                 ~0.307$\pm$~0.119 &                          ~0.860$\pm$~0.098 &   ~9.111$_{-0.430}^{+0.438}$ \\
  J012217+052044 &                       ~0.594$\pm$~0.069 &                                 ~0.467$\pm$~0.100 &                          ~0.417$\pm$~0.096 &   ~8.762$_{-0.423}^{+0.448}$ \\
  J012910+145935 &                       ~0.193$\pm$~0.017 &                                 ~0.719$\pm$~0.091 &                          ~0.965$\pm$~0.091 &   ~9.154$_{-0.305}^{+0.583}$
\enddata
\end{deluxetable*}

\startlongtable
\begin{deluxetable*}{lcrrrcc}
\tablecaption{Measurements of the LyC. A full version of this table is available online.\label{tab:flyc}}
\tablewidth{\textwidth}
\tablehead{
\colhead{Object} & \colhead{$\lambda_{LyC}$ \tablenotemark{a}} & \colhead{Dark} & \colhead{Sky} & \colhead{Source} & \colhead{$P(>N|B)$} & \colhead{$f_{LyC}\times10^{17}$ \tablenotemark{b}}
\\
\colhead{} & \colhead{\AA} & \colhead{counts} & \colhead{counts} & \colhead{counts} & \colhead{} & \colhead{erg s$^{-1}$ cm$^{-2}$ \AA$^{-1}$}
}
% 1.30E-07
% 2.28E-06
% 3.04E-07
\startdata
J003601+003307 &   860 &  62.388 &  19.298 &    13.314 &                0.066 &                  $<$~0.114 \\
J004743+015440 &   860 &  12.438 &  ~6.664 &    25.898 & $1.305\times10^{-7}$ & ~1.557$_{-0.407}^{+0.457}$ \\
J011309+000223 &   890 &  ~8.970 &  ~6.471 &    20.559 & $2.282\times10^{-6}$ & ~1.445$_{-0.428}^{+0.490}$ \\
J012217+052044 &   850 &  34.209 &  ~5.945 &    34.845 & $3.045\times10^{-7}$ & ~1.118$_{-0.266}^{+0.292}$ \\
J012910+145935 &   890 &  11.290 &  ~9.734 &     2.237 &                0.266 &                  $<$~0.556
\enddata
\tablenotetext{a}{Rest-frame central wavelength of 20 \AA\ LyC spectral window}
\tablenotetext{b}{LyC flux density corrected for MW extinction}
%\tablenotetext{c}{~\citet{2016Natur.529..178I}}
%\tablenotetext{d}{~\citet{2016MNRAS.461.3683I}}
%\tablenotetext{e}{~\citet{2018MNRAS.474.4514I}}
%\tablenotetext{f}{~\citet{2018MNRAS.478.4851I}}
%\tablenotetext{g}{~\citet{2019ApJ...885...57W}}
\end{deluxetable*}

\startlongtable
\begin{deluxetable*}{lccc}
\tablecaption{Empirical \fesclycrel\ and absolute \fesclyc\ derived from {\sc Starburst99} using continuous star formation predictions and H$\beta$ or the burst predictions fit to the {\it HST}/COS spectrum for the combined LzLCS and \pubsamp\ samples. A full version of this table is available online.\label{tab:fesc}}
\tablewidth{\textwidth}
\tablehead{
\colhead{Object} & \colhead{$f_{LyC}/f_{1100}$} & \colhead{$f_{esc}^{LyC}$(H$\beta$)} & \colhead{$f_{esc}^{LyC}$(UV)}
}
\startdata
  J003601+003307 &                    $<$~0.017 &                           $<$~0.005 &                     $<$~0.010 \\
  J004743+015440 &   ~0.040$_{-0.010}^{+0.012}$ &          ~0.049$_{-0.012}^{+0.014}$ &    ~0.015$_{-0.004}^{+0.019}$ \\
  J011309+000223 &   ~0.040$_{-0.012}^{+0.013}$ &          ~0.123$_{-0.035}^{+0.042}$ &    ~0.020$_{-0.012}^{+0.019}$ \\
  J012217+052044 &   ~0.048$_{-0.012}^{+0.013}$ &          ~0.084$_{-0.019}^{+0.022}$ &    ~0.038$_{-0.014}^{+0.057}$ \\
  J012910+145935 &                    $<$~0.011 &                           $<$~0.014 &                     $<$~0.007 \\
\enddata
%\tablenotetext{a}{\citet{2016Natur.529..178I}}
%\tablenotetext{b}{\citet{2016MNRAS.461.3683I}}
%\tablenotetext{c}{\citet{2018MNRAS.474.4514I}}
%\tablenotetext{d}{\citet{2018MNRAS.478.4851I}}
%\tablenotetext{e}{\citet{2019ApJ...885...57W}
\end{deluxetable*}

\end{document}